\documentclass[journal,onecolumn,12pt]{IEEEtran}

\usepackage{cite}   
\usepackage{amsmath,amssymb}
\usepackage{enumerate}
\usepackage{array}
\newtheorem{definition}{Definition}
\newtheorem{theorem}{Theorem}

\newtheorem{proposition}{Proposition}

\usepackage{graphicx}
\usepackage{epsfig}

\begin{document}

\title{Permutation Polynomial Interleavers: An Algebraic-Geometric Perspective}    

\author{Oscar Y. Takeshita  \\
Dept. of Electrical and Computer Engineering \\
2015 Neil Avenue \\
The Ohio State University \\
Columbus, OH 43210 \\ 
Takeshita.3@osu.edu\\
\vspace{2em}
Submitted to the IEEE Transactions on Information Theory\\
January 12, 2006\\
}
\date{\today}

\maketitle

\begin{abstract}
  An interleaver is a critical component for the channel coding
  performance of turbo codes.  Algebraic constructions are 
  important because they admit analytical designs and
  simple, practical hardware implementation. The spread factor of an
  interleaver is a common measure for turbo coding
  applications. Maximum-spread interleavers are interleavers whose
  spread factors achieve the upper bound. An infinite sequence of
  quadratic permutation polynomials over integer rings that generate
  maximum-spread interleavers is presented. New properties of
  permutation polynomial interleavers are investigated from an
  algebraic-geometric perspective resulting in a new non-linearity metric
  for interleavers. A new interleaver metric that is a function of both
  the non-linearity metric and the spread factor is proposed.  
  It is numerically demonstrated that the spread factor has a
  diminishing importance with the block length. A table of good
  interleavers for a variety of interleaver lengths according to the
  new metric is listed. Extensive computer simulation results with impressive
  frame error rates confirm the efficacy of the new metric. Further,
  when tail-biting constituent codes are used, the resulting turbo
  codes are quasi-cyclic.    
\end{abstract}


\pagebreak

\section{Introduction}

Interleavers for turbo codes~\cite{sun:tak:pp, nim:isit04, berrou:int,
  Bravo, giulietti, Sadjadpour, tak:cos:it, crozier2,crozier_DRP,
  danesh:it,tak:cos:linear, dolinar,barbulescu, Berrou} have  
been extensively investigated. However, the design of interleavers for
turbo codes is complex enough and we believe there are still several
relevant open questions. Recently, Sun and Takeshita~\cite{sun:tak:pp}
suggested the algebraic approach of using permutation polynomial-based
interleavers over integer rings. In this paper, we expand the theory 
in~\cite{sun:tak:pp} by adding a geometric and group-theoretic
perspective. Several new results and important and interesting new
questions arise with this framework. 

We believe that the approach in~\cite{sun:tak:pp} has
major advantages over any earlier interleaver constructions
(either pseudo-random or structured) because it simultaneously provides:
\begin{itemize}
\item Excellent error performance with practical
  code-lengths~\cite{tak:mcf}.
\item Completely algebraic structure with elegant and relevant
  properties~\cite{tak:mcf,ryu:tak:qinv}. 
\item Efficient implementation~\cite{jpl2} with high-speed, low-power
  consumption, and little memory requirements.
\end{itemize}

\subsection{Interleavers and Permutation Polynomials}

An interleaver is a device that permutes a sequence of
$N$ symbols. Let each symbol be indexed by an element in the set
$S_N=\{0,1,\ldots, N-1\}$. Then the interleaver can be represented by
a one-to-one onto (permutation) function $f:S_N\rightarrow S_N$ given
by $f:x\mapsto f(x)$.  

Permutation polynomials over $\mathbb Z_N$ are functions that belong
to the ring ${\cal R}_N=\mathbb Z_N[x]$, i.e., polynomials of the
form $q(x)=\sum_{i=0}^K q_i x^i\pmod{N}$ such that $q_i\in
\mathbb{Z}_N$ and  $q:\mathbb Z_N\rightarrow \mathbb Z_N$ is a permutation
function. In earlier work, we only treated {\em constant-free} PPs, i.e., $q_0=0$; in this paper,
we also consider $q_0\neq 0$.  
The set of permutation functions (not necessarily permutation
polynomials) for a given interleaver length $N$ will be denoted by
${\cal P}_N$. The set ${\cal P}_N$ is no longer a
ring but its elements form a group under function composition. 

Permutation polynomials over $\mathbb{Z}_N$ naturally generate 
interleavers where $S_N$ gets equipped with a finite ring
structure $\mathbb{Z}_N$, which is not necessary for the definition of
an arbitrary interleaver but allows a useful algebraic
treatment of PP interleavers. Replacing $S_N$ by
$\mathbb{Z}_N$ in the case of arbitrary interleavers brings no loss of
  generality, therefore from here on we will always use
  $\mathbb{Z}_N$. The algebraic structure of PPs
  over finite integer rings has
already been proven very successful in producing  turbo
codes~\cite{tak:mcf} and low density parity check (LDPC) 
codes~\cite{tak:qpp_ldpc} with excellent error performance 
compared with the best known constructions and similar parameters.

Conditions for the coefficients of a polynomial over $\mathbb Z_N$  to
be a PP for an arbitrary integer $N$ were  studied
in~\cite{sun:tak:pp}. However, the conditions are not the simplest for
polynomials of degrees larger than two; to the best of our knowledge,
simple necessary and sufficient conditions for arbitrary degrees are
only known when $N$ is a power of two~\cite{rivest}. If $q(x)$ is of
second degree and $N$ is arbitrary then a simple necessary and
sufficient condition was proved in~\cite{sun:tak:pp} and summarized
in~\cite{ryu:tak:qinv,tak:mcf}. Interleavers
constructed using second degree or quadratic permutation polynomials
will be called QPP interleavers, linear interleavers if the
PP is of first degree, and PP interleavers for PPs of arbitrary
degrees.  

\subsection{Good Interleaver Measures}

Because the number of distinct permutations of length $N$ is $N!$, it is very
important that effective interleaver measures for turbo codes are
defined; this considerably reduces the number of interleavers that 
still need to be filtered by costly analysis and computer simulations
for a complex turbo codec system. 
The main drawback of the theory in~\cite{sun:tak:pp} was that despite it
provided some rules for choosing good permutation polynomials (PP), the
procedures were too complex when handling input weights larger than
two. In this paper, we propose a new simple but 
effective measure $\Omega$ for interleavers. Two typical measures for
interleavers in turbo coding are the spread factor $D$~\cite{dolinar,crozier2} 
and ``randomness.'' In this paper, the notion of
``randomness'' is replaced by a more principled concept of a degree of
non-linearity $\zeta$ of an interleaver. An interleaver is represented
by what we call an {\em interleaver-code}, which is the geometric
representation of an interleaver by pairs of coordinates $(x,f(x))$
forming {\em points} in $\mathbb{Z}_N^2$. The degree of non-linearity 
$\zeta$ measures the number of disjoint orbits (a set of points) of
the action of an isometry group of the interleaver-code. The new 
measure is simply the product of the logarithm of the spread factor by
the new non-linearity metric, i.e., $\Omega=\ln(D)\zeta$. The
algebraic-geometric nature of PPs allows a very
efficient selection of PPs that maximizes the
new metric. 

DRP interleavers~\cite{crozier_DRP} are among the best
known interleavers for turbo codes with a combined excellent error
rate performance (exceeding that of $S-$random
interleavers~\cite{dolinar}) and simplicity but are not fully
algebraic. The efficacy of the new metric is shown by simulation
curves of several turbo codes using PP interleavers with impressive
frame error rate performance, similar to the ones with DRP
interleavers. However, the selection of good PP interleavers 
(particularly QPP interleavers) is much simpler with the results presented
this paper.  

\subsection{The Relationship between $D$ and $\zeta$}

The non-linearity metric $\zeta$ is shown to be inversely
related to the degree of shift-invariance $\epsilon=1/\zeta$ of an
interleaver. For QPP interleavers, the shift-invariance $\epsilon$ is
computed in closed form as function of the second degree coefficient,
which gives us a complete control of this parameter. When tail-biting
codes are used as constituent codes, turbo codes using QPP
interleavers become quasi-cyclic; for those codes, it is predicted
that the multiplicity of many low-weight codewords is typically
multiples of the shift-invariance $\epsilon$. It is also shown that 
a lower bound on the spread factor $D$ constrained to points within an 
orbit of an interleaver-code has a closed form for QPP
interleavers. An immediate tie between $D$ and $\zeta$ is then
established for QPP interleavers where one would need to trade for
either larger $D$ or $\zeta$.

This paper is organized as follows. In Section~\ref{sec:ms},
maximum-spread interleavers are investigated and a list of good
permutation polynomials for turbo codes for several 
lengths that maximizes the spread factor $D$ is found in
Table~\ref{tab:maxqpp}. A new metric for interleavers is developed in
Section~\ref{sec:qpp_nl_si}; we provide a table
(Table~\ref{tab:maxomega}) of good permutation polynomials for turbo
codes according to the new metric. Numerical results are shown in
Section~\ref{sec:nr} via the computation of the distance spectra and
computer simulations for several codes. Finally,  
conclusions and possible future directions are discussed in
Section~\ref{sec:conclusions}.

\section{Maximum-Spread Interleavers}
\label{sec:ms}

In this section, the spread factor of an interleaver is revisited. An
infinite sequence of QPP interleavers achieving the upper bound on
the spread factor is presented. Further, a list of QPP interleavers with
optimal spread factors for several interleaver lengths reported in the
literature is provided.  

\subsection{A Geometric View of Interleavers}

\begin{figure}[htbp]
  \centering
  \includegraphics[width=0.30\hsize,angle=270]{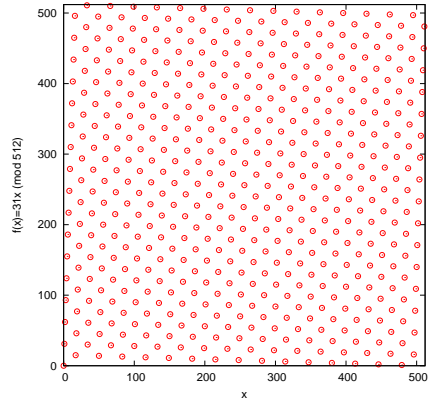}
  \caption{The interleaver $l(x)=31x\pmod{512}$ viewed as an interleaver-code
    $L=\Phi(l)$ over $\mathbb Z_N^2$.}
  \label{fig:MS512_0_31}
\end{figure}

In the algebraic-geometric treatment in this paper, it will be
convenient to view an interleaver represented by a permutation
function $f(x)$ (not necessarily a permutation polynomial) as an
interleaver-code $F$ 
(not to be confused with a turbo code) under the natural mapping
$\Phi:{\cal P}_N\rightarrow \mathbb{Z}_N^2$ given by $\Phi:f(x)\mapsto
F$, where $F=\{(x,f(x))|x\in \mathbb{Z}_N\}$. We will call a
pair $(x,f(x))$ as a {\em point} $p_x=(x,f(x)) \in F$. Let a 
linear interleaver be given by the PP $l(x)=31x\pmod{512}$; the corresponding
interleaver-code $L=\Phi(l)$ is plotted in~Fig.~\ref{fig:MS512_0_31}.

The notion of {\em distance} or {\em metric} between points in an interleaver-code
$F$ will be of central role. We will mainly use two different metrics.

\subsubsection{$L_1$ metric}
Let us define a distance $\delta$ in $\mathbb Z_N^2$ to
form a metric space $(\mathbb Z_N^2,\delta)$,

\[
\delta(p_{x_1},p_{x_2})=|x_1-x_2|+|f(x_1)-f(x_2)|,
\]

i.e., the metric is the $L_1$ or Manhattan metric.

\subsubsection{Lee metric}
Let us define a distance $\delta_N$ in $\mathbb Z_N^2$ to
form a metric space $(\mathbb Z_N^2,\delta_N)$,

\[
\delta_N(p_{x_1},p_{x_2})=|x_1-x_2|_N+|f(x_1)-f(x_2)|_N,
\]

\noindent where 
\[
|i-j|_N=\min\{(i-j)\pmod{N},(j-i)\pmod{N}\}
\]
is the Lee distance~\cite{lee} between $i$ and $j$ modulo
$N$. Therefore, $\delta_N$ is a two-dimensional Lee metric.

\subsection{The Spread Factor of an Interleaver}

 The spread factor~\cite{dolinar} of an interleaver is a
popular measure of merit in turbo coding applications. The spread
factor of an interleaver $f(x)$ over the metric space
$(\mathbb{Z}_N^2,\delta)$ is defined as

\begin{equation}
D_E(f)=\min_{\substack{i,j\in \mathbb{Z}_N\\i\neq
    j}}\{\delta(p_i,p_j)|p_i,p_j\in F\}.
\label{eq:spread}
\end{equation}

\noindent
The root of this
measure is the early recognition in turbo coding that self-terminating
information sequences of weight two create low-weight
codewords~\cite{dolinar}. Divsalar and Dolinar then proposed a construction of
linear interleavers achieving spread factors $D_E$ equal to or close to
$\sqrt{2N}$. They have concluded that by using linear interleavers, the
minimum distance of turbo codes conditioned to weight-two
self-terminating information sequences grows roughly as
$\sqrt{2N}$. However, recent results show that the true 
minimum distance grows asymptotically only at most
logarithmically~\cite{breiling} for all interleavers. 
If we use the metric space $(\mathbb{Z}_N^2,\delta_N)$ 
then the following definition of spread factor is also
possible~\cite{crozier2}: 

\[
D(f)=\min_{\substack{i,j\in \mathbb{Z}_N\\i\neq j}}\{\delta_N(p_i,p_j)|p_i,p_j\in F\}.
\]

\noindent
The spread factor $D$ is mathematically more convenient because 
the metric space $(\mathbb{Z}_N^2,\delta_N)$ is ``isotropic'' (in the
sense of having no boundaries). It is also more suitable when
designing turbo codes with tail-biting convolutional constituent
codes~\cite{crozier2}. An analytical   
proof on the upper bound $ub_D(N)$ of $D$ was recently shown to be 
$\sqrt{2N}$~\cite{boutillon}. Clearly $D\leq D_E$ by the definition of
$\delta$. We show $ub_{D_E}(N)$ is close to
$ub_D(N)$ by computing a new upper bound
$ub_{D_E}(N)$. It is a constructive bound for certain values of $N$. See 
Appendix~\ref{sec:bound} for a sketch of a proof.  

\begin{equation}
ub_{D_E}(N)=
\left\{
\begin{array}{cl}
\frac{2(N-1)}{\sqrt{2N}-1}, & N=2p^2, \quad p=2,3,4\ldots\\
\frac{2(N-1)}{\sqrt{2N-1}-1}, & N=p^2+(p-1)^2, \quad p=2,3,4,\ldots. \\
\end{array}
\right.
\label{eq:debound}
\end{equation}

\begin{figure}[htbp]
  \centering
  \includegraphics[width=0.6\hsize]{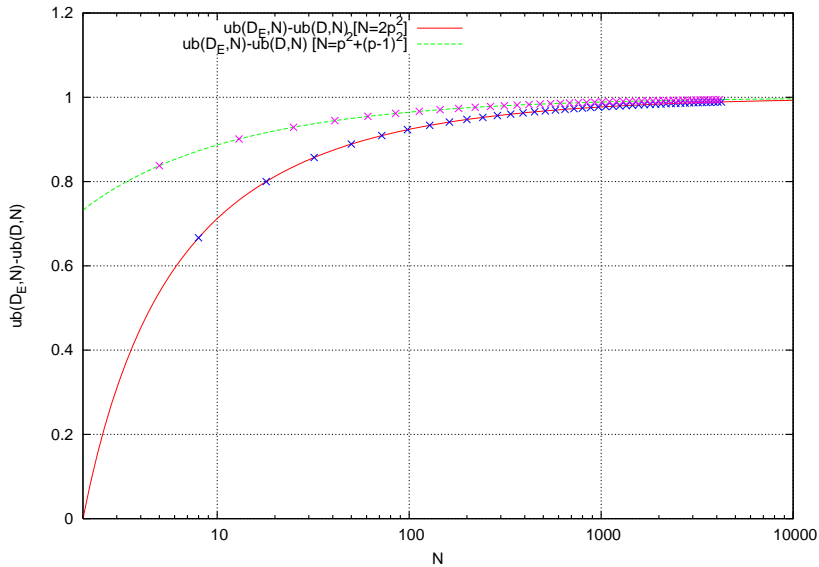}
  \caption{The difference $ub_{D_E}(N)-ub_D(N)$.}
  \label{fig:ubounds}
\end{figure}

\noindent
The difference $ub_{D_E}(N)-ub_D(N)$ goes to
1 as $N$ grows (See Fig.~\ref{fig:ubounds}).

\noindent 
Only one example was found (for  a limited search) of an interleaver
(up to symmetries) that has a  
spread factor $D_E$ exceeding $ub_D(4)$. For $N=4$, the interleaver
$t:\mathbb{Z}_4\rightarrow \mathbb{Z}_4$ defined by $t(0)=1$,
$t(1)=3$, $t(2)=0$, and $t(3)=2$ has  

\[
ub_D(4)=2.8284<D_E(t)=3. 
\]

\noindent
Our new bound (\ref{eq:debound}) is not defined at $N=4$ but for
this case clearly an upper bound is $ub_{D_E}(4)=3$ by inspection, i.e.,
an upper bound on $D_E$ strictly larger than $ub_D(4)$ is achievable. 
The main message learned is that in practice considering $D$ is good enough. For
the remaining of the paper, we will only use $D$ instead 
of $D_E$, unless otherwise noted, because $D$ lets us use the algebra
of $\mathbb{Z}_N$.

\begin{definition} An interleaver of length $N$ is a {\em
    maximum-spread} interleaver if it achieves a spread factor $D$
  equal to the upper bound\footnote{Because the spread factor must be
    an integer, the upper bound is straightforwardly tightened to $\lfloor
    \sqrt{2N}\rfloor$. However, this will be of little relevance in
    this paper and we keep $\sqrt{2N}$ for simplicity, unless
    otherwise noted.} $\sqrt{2N}$. 
\end{definition}

Once more, Dolinar and Divsalar~\cite{dolinar} have reported that
linear interleavers either achieve or closely approximate a
spread factor of $\sqrt{2N}$ for any $N$. In particular, if $N$ is
twice of a perfect square

\begin{equation}
N=2n^2,\quad n=1,2,3,\ldots,
\label{eq:2perfsq}
\end{equation}

\noindent
they have given {\em all} maximum-spread linear interleavers
of the form $f(x)=f_1x\pmod{N=2n^2}$ by an explicit
simple condition for $f_1$.
 However, they pointed out that the resulting interleavers were not
 good for turbo coding because of their high regularity (see
 Fig.~\ref{fig:MS512_0_31}, which corresponds to a plot of a
 maximum-spread linear interleaver).  This fact is also theoretically
 addressed by the linear interleaver asymptote in~\cite{tak:cos:it}, which implied
the existence of low-weight codewords of input-weight four and a high
multiplicity, close to $N$. Therefore they
proposed a semi-random interleaver construction
algorithm~\cite{dolinar}, generating the so-called $S$-random 
interleavers with a parameter $S$, but the algorithm sacrificed the spread
factor $D$. The obtained spread factor is typically  $D=S+1\leq \sqrt{N/2}+1$,
i.e., smaller than about 50\% of the upper bound
$ub_D(N)=\sqrt{2N}$. An $S$-random interleaver yield turbo codes with very good
error rate performances and became the typical benchmark
interleaver. One of the main drawbacks of 
$S$-random interleavers is a costly storage of a
sequence of $N$ integers needed to specify the interleaver. Because the
construction algorithm relies heavily on a pseudo-random number
generator, the sequence has little margin for compression. This
characteristic also makes more difficult  for an accurate reproduction
of the results by others because typically only the parameter $S$ is
reported in the literature; however, the repeatability problem is not
so critical because in general, for a given parameter $S$, instances of $S$-random
interleavers perform similarly for error rates that are not extremely
low, which also reflects a good minimum distance of the associated turbo
code. Crozier proposed two interleaver constructions~\cite{crozier2}
that attempt maximization of the spread factor $D$ but avoid or
minimize the regularity of linear interleavers: the high-spread
construction and the dithered-diagonal construction. The
dithered-diagonal interleavers~\cite{crozier2} are reported to be 
maximum-spread for interleaver sizes as in (\ref{eq:2perfsq})
and have large spread factors for others. Dithered-diagonal
interleavers require $n=\sqrt{N/2}$ integer parameters for their definition. An
impressive error performance, exceeding the performance of $S$-random
interleavers was shown for $N=512$~\cite{crozier2}. This was a great 
progress in the field of interleaver constructions for turbo codes because it
produced a large spread factor $D$ combined with sufficient
irregularity to produce an excellent error performance and a
much smaller number of integer parameters specifying the interleaver
compared with $S$-random interleavers. Subsequently, Crozier and
Guinard proposed dithered relatively prime (DRP)
interleavers~\cite{crozier_DRP}. DRP interleavers also perform very 
well but require a much smaller number of integer parameters for their
specification. Their typical choice for a good performance
requires about 10 integers\footnote{The length of the input parameters
  are more precisely defined in Section~\ref{sec:drp_vs_qpp}.} for
their dither parameters $M=R=4$. For an 
excellent performance, their dither parameters $M=R=8$ require about 18
integers.

\subsection{The Spread Factor of QPP Interleavers}

\begin{figure}[htbp]
  \centering
  \includegraphics[width=0.7\hsize]{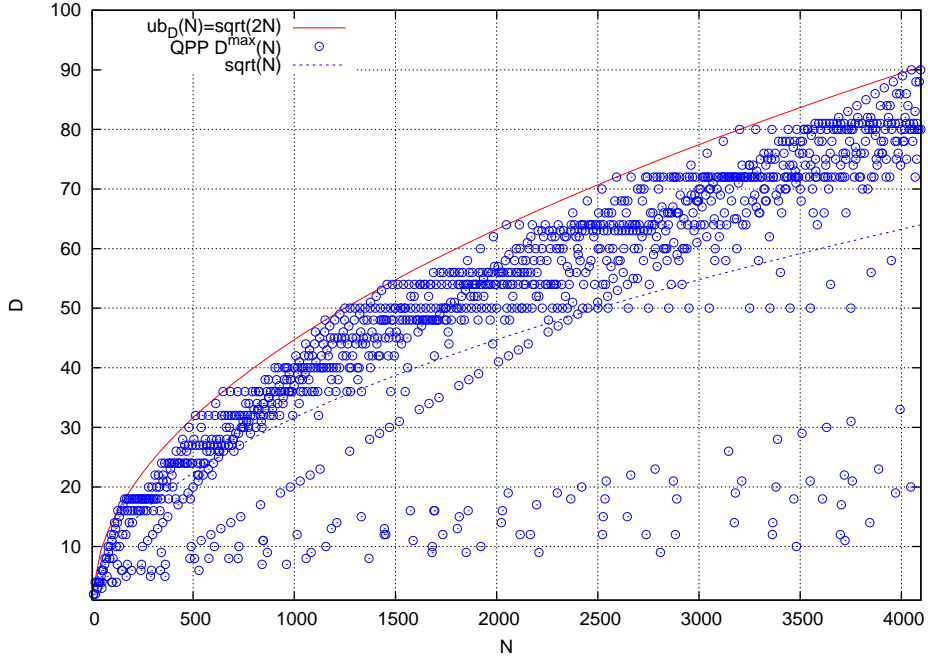}
  \caption{Maximum achievable spreads $D^{\max}(N)$ with QPP interleavers for a
    given $N$.}
  \label{fig:dmax}
\end{figure}

Interleavers based on QPP require only 2 integers as parameters with
the simplicity of only a simple quadratic congruence as the algorithm
to generate the permutation sequence. In fact, QPP interleavers can be implemented
in hardware with only additions and comparisons~\cite{jpl2}.
If we fix $N$, what is the largest spread for a QPP interleaver? An
exhaustive search for the largest achievable spread $D^{\max}(N)$ for
$2\leq N \leq 4096$ is shown in Fig.~\ref{fig:dmax}. Due to several
algebraic and geometric properties of QPP interleavers explained in
Section~\ref{sec:qpp_nl_si}, the exhaustive search is efficiently completed
in a very short time on a regular personal computer using
Theorem~\ref{th:qpp_search}.  

\noindent
A few of the polynomials for some commonly found interleaver lengths
in the literature are reported in Table~\ref{tab:maxqpp}. Some of the
QPPs in Table~\ref{tab:maxqpp} 
generate very good turbo codes, however, better QPPs that do not
simply attempt maximization of the spread factor are provided in
Section~\ref{sec:new_metric}. 

\begin{table}[htbp]
\centering
\caption{QPPs with the largest spread factors $D^{\max}(N)$.}
\begin{tabular}{|r|c|c|c|c|c|} \hline
$N$  & $f(x)$ & $f^{-1}(x)$ & $D^{\max}(N)$ &$\zeta$ & $\zeta^\prime$ \\ \hline
40   &  $x+10x^2$ &  $21x+10x^2$ & 4 & 2 & 2 \\
80   &  $9x+20x^2$  & $49x+20x^2$&10 & 2 & 2 \\
128  &  $15x+32x^2$  & $111x+32^2$&16& 2 & 2 \\
160  &  $19x+40x^2$  & $59x+40x^2$&16& 2 & 2 \\
256  & $15x+32x^2$  & $239x+32x^2$&16& 4 & 3 \\
320  & $19x+40x^2$  & $59x+40x^2$&20& 4 & 3 \\
400  & $17x+100x^2$  & $153x+100x^2$ &20& 2 & 2\\
408  & $25x+102x^2$  & $253x+102x^2$ &24& 2 & 2\\
512  & $31x+64x^2$  & $479x+64x^2$ &32  & 4 & 3 \\
640  & $39x+ 80x^2$  & $279x+80x^2$ &32 & 4 & 3 \\
752  & $31x+ 188x^2$  & $655x+188x^2$ &32 & 2 & 2 \\
800  & $17x+ 80x^2$  & $753x+ 240x^2$ & 32  & 5 & 5 \\
1024 & $123x+256x^2$ & $691x+256x^2$ &34& 2 & 2 \\
1280 & $39x+80x^2$   & $279x+80x^2$ &40& 8 & 4 \\
1504 & $183x+376x^2$  & $263x+376x^2$ &46& 2 & 2 \\
1600 & $49x+100x^2$   & $849x+700x^2$ &50& 8 & 4 \\
2048 & $63x+ 128x^2$  & $1983x+128x^2$ &64& 8 & 4 \\
2560 & $79x+160x^2$  & $1199x+160x^2$  &64& 8 & 4 \\
3200 & $79x+800x^2$  & $3119x+800x^2$ &80 & 2 & 2 \\
4096 & $173x+1024x^2$ & $2853x+1024x^2$ &80& 2 & 2\\
\hline
\end{tabular}
\label{tab:maxqpp}
\end{table}

Observing once more Fig.~\ref{fig:dmax}, 
at least in the searched range, the fraction of maximum-spread
QPP interleavers is very small. The same plot, however, reveals that the
majority of QPP interleavers with the largest spreads lie between
$ub_D(N)=\sqrt{2N}$ and $\sqrt{N}$ (about 70\% of the upper
bound). Recall that $S$-random interleavers typically achieve only 50\%
of $ub_D(N)$. It is also important that spread factors $D^{\max}(N)$ are shown only 
when there exist QPP interleavers for a given $N$ in a strict sense,
i.e., with irreducible degree, as explained in
Section~\ref{sec:ms_qpp}. Linear interleavers always exist for any $N$
but this 
is not true for QPP  
interleavers~\cite{sun:tak:pp,ryu:tak:qinv}. Between $2\leq N\leq
4096$, there are 1190 values of $N$ that produce a QPP interleaver
(roughly 29\%). This raises another question as to what values of $N$
have QPP interleavers in the strict sense.

\begin{theorem} Let $N=2^3m=8m$ for some $m\in \{1,2,3,\ldots\}$. Then
  there exists a quadratic permutation polynomial
  $f(x)=f_1x+f_2x^2\pmod{N}$ with irreducible degree. 
\label{th:multp8}
\end{theorem}

\begin{proof}
One simply chooses $f_2=2m$ and $f_1=1$, which satisfy the necessary and
sufficient conditions for a quadratic permutation polynomial
in~\cite{sun:tak:pp}. The chosen polynomial is not reducible to a
linear polynomial because, by Theorem~\ref{th:qpp_nl_degree}, its degree
of non-linearity is $\zeta=2$.
\end{proof}

\noindent
Theorem~\ref{th:multp8} may be of practical relevance because it guarantees the
existence of QPP interleavers for all positive multiples of a typical
computer byte size of 8.  

\subsection{The Maximization of the Spread Factor $D$}
\label{sec:Dmaximization}

The maximization of the spread factor $D$ is clearly beneficial in the
minimization of the number of low-weight codewords caused by
self-terminating weight-2 input sequences. This concept trivially
generalizes to short bursts of self-terminating input
sequences, which then include the classical weight-2 input. Should we
then look for interleavers that always maximize the spread factor? To
understand the context of this question, we make the following
observations: 

\begin{itemize}
\item Many linear interleavers are maximum-spread but suffer from 
high-multiplicity low-weight codewords~\cite{tak:cos:it}. 
\item At least one dithered diagonal interleaver~\cite{crozier2}
  is maximum-spread and provides an impressive error performance. 
\item DRP interleavers which maximize error performance are typically
  not maximum-spread~\cite{crozier_DRP}.
\end{itemize}

Although it is not possible to generalize from the single example
reported in~\cite{crozier2}, from observation 2), the best
interleavers from an error rate perspective may be the ones that either
achieve or closely approximate a maximum-spread interleaver and
simultaneously have a large degree of ``randomness.'' Interleavers with
some structure such as DRP and our PP interleavers (or low-entropy, as
discussed in Section~\ref{sec:drp_vs_qpp}) may need to give up some spread
factor in order to convert some of its entropy for ``randomness.'' 

A common practice for a more accurate estimation of
error performance away from asymptotics is not only to examine the
minimum distance of a code but also the distance profile. Similarly, as
we better understand PP interleavers, the spread profile may become more
important. We define the multiplicities of the {\em spread profile} of an
interleaver-code $F$ as 
\[
A_{D_i,p_x}=|\{p_{x_1}| i=\delta_N(p_x,p_{x_1}),
p_x\neq p_{x_1},p_x,p_{x_1}\in F\}|,
\]

\noindent
where $D_i,\quad i=1,2,\ldots,\lfloor \sqrt{2N}\rfloor$, is the $i$-th
spread. Good interleaver-codes are in general 
non-linear (see Section~\ref{sec:qpp_nl_si}), and therefore the spread
profile above is a function of each point $p_x\in F$. 
For arbitrary interleavers, computing the spread profile becomes
difficult without a brute force search because the interleaver-code induced by
$\Phi$ is highly non-linear. However, for PP interleavers, the spread
profile can be inspected just for the representatives of the orbits. 
In fact, what we proposed in~\cite{sun:tak:pp} can be interpreted 
as the creation of ``spectral nulls'' in the spread profile for
self-terminating weight-2$m$ sequences and giving up the maximization
of the first spread spectral line $D_1$. The shaping of the spread
profile may be a promising way for designing and searching for good
PPs.  

\subsection{An Infinite Sequence of Maximum-Spread QPP Interleavers}
\label{sec:ms_qpp}

\begin{theorem}
The following is an infinite sequence of QPPs that generate {\em
  maximum-spread} interleavers. 
\begin{equation}
f(x)=(2^k-1)x+2^{k+1}x^2 \pmod{2^{2k-1}}\quad k=1,2,3,\ldots
\label{eq:sequence}
\end{equation}
\label{th:sequence}
\end{theorem}
\begin{proof}
See Appendix~\ref{sec:sequence}.
\end{proof}

Strictly, we have  QPP interleavers only when $k>3$. The first observation
is that for $k=1$ and $k=2$, the corresponding QPPs $f(x)$ are immediately
reduced to first degree polynomials because the second degree coefficient
$f_2\equiv 0 {\pmod N}$. We now show after some preliminaries
that for $k=3$, the QPP $f(x)$ is also reducible
to a first degree polynomial although $f_2=2^{k+1}=16\not \equiv 0\pmod{N=32}$. 

\begin{definition}
A polynomial $z(x) \pmod{N}$ that evaluates to zero for all $x$, i.e.,
$z(x)\equiv 0 \pmod{N} \quad \forall x$ is called a {\em zero-polynomial}.
\end{definition}

\begin{proposition} Let $N$ be an integer factorable as $N=pq$.
The following is a zero-polynomial of degree $p$.
\[
z(x)=mq\prod_{i=0}^{p-1} (x+k+i) \pmod {N}\quad \forall k,m\in \mathbb{Z}_N, 
\]
\label{prop:zero}
\end{proposition}
\begin{proof} Exactly one of the numbers in the sequence
$x+k+i,0\leq i<p$ is congruent to 0 modulo
$p$. Therefore $z(x)$ must evaluate to zero for all $x$ because
$mqp\equiv 0\pmod{N}$. 
\end{proof}

\begin{proposition} Let a polynomial $p(x) \mod{N}$ and a
  zero-polynomial $z(x)\mod{N}$. Then $s(x)\equiv p(x)+z(x)\pmod{N}$,
  i.e., $s(x)$ and $p(x)+z(x)$ are equivalent functions modulo $N$.
\end{proposition}
\begin{proof}
This follows directly from the definition of a zero-polynomial.
\end{proof}

From Proposition~\ref{prop:zero}, the following is a zero-polynomial
of second degree for $N=32$

\[
z(x)=16x(x+1)=16x^2+16x\pmod{32}.
\]

Therefore for $k=3$, by adding $f(x)$ to $z(x)$ we obtain the equivalent
first degree polynomial 

\[
s(x)\equiv f(x)+z(x)=(16x^2+7x)+(16x^2+16x)\equiv 23x\pmod{32}.
\]

For $k>3$, the polynomials are not reducible to first degree
polynomials because they have a degree of non-linearity $\zeta$ larger
than 1 as explained in Section~\ref{sec:qpp_nl_si} (In fact, for $k=1,2,3$
we have $\zeta=1$).
The first six terms of maximum-spread QPPs that are not
reducible to first degree polynomials are shown in
Table~\ref{tab:msinterleavers}. The last three columns of the table
are the degrees of non-linearity $\zeta$, the refined degree of
non-linearity $\zeta^\prime$ and the degree of shift-invariance $\epsilon$
as explained in Section~\ref{sec:qpp_nl_si}.  
\begin{table}[htbp]
\centering
\caption{Examples of Maximum-Spread $QPP$ interleavers}
\begin{tabular}{|c|c|c|c|c|c|c|c|c|c|}\hline
$k$& $N$ & $f(x)$ & $f^{-1}(x)$ & $D=ub_D(N)$ & $\zeta$ & $\zeta^\prime$
& $\epsilon$\\ \hline
 4 & 128 & $15x+32x^2$ &  $-17x+32x^2$ & 16 & 2 & 2 & 64 \\
 5 & 512 & $31x+64x^2$ & $-33x+64x^2$ & 32 & 4 & 3 & 128 \\
 6 & 2048 & $63x+128x^2$ & $-65x+128x^2$ & 64 & 8 & 4  & 256 \\
 7 & 8192 & $127x+256x^2$ & $-129x+256x^2$ & 128 & 16 & 7 & 512\\
 8 & 32768 & $255x+512x^2$ & $-257x+512x^2$ & 256 & 32 & 12 & 1024\\  
 9 & 131072 & $511x+1024x^2$ & $-513x+1024x^2$ & 512 & 64 & 23 & 2048\\ \hline 
\end{tabular}
\label{tab:msinterleavers}
\end{table}

The inverse functions $f^{-1}(x)$ are also provided in
Table~\ref{tab:msinterleavers}. The closed form expression for $f^{-1}(x)$ is 

\[
f^{-1}(x)=(-2^k-1)x+2^{k+1}x^2\pmod{2^{2k-1}}.
\]

One easily verifies that $f(f^{-1}(x))\equiv f^{-1}(f(x))\equiv x \pmod{N}$. 
For general QPPs, we are not aware of a closed form expression for the
inverse functions. Further, not all QPPs have an inverse polynomial
that is a QPP. This was first observed in~\cite{jplppnotes}. However,
if it exists, it is efficiently computed algebraically using
the extended Euclidean algorithm~\cite{ryu:tak:qinv}. It is easily
verified that the necessary and sufficient condition for the existence
of a QPP inverse~\cite{ryu:tak:qinv} for the polynomials in
Theorem~\ref{th:sequence} is satisfied.

\section{An Algebraic-Geometric View of Interleavers}
\label{sec:qpp_nl_si}

In this section, PP interleavers are tied to an algebraic-geometric view
by examining the isometries of the associated interleaver-codes. A new
measure for interleavers arises as a consequence. The measure is
easily computed for QPP interleavers using their algebraic-geometric
properties. A list of QPP interleavers which maximize the new measure
is provided. Finally, some comments on larger degree PPs are made.

\subsection{A New Measure for Interleavers}

The following framework is well known in the 
context of geometrically uniform codes~\cite{forney_gu}. A treatment
of groups and geometry is found in~\cite{neumann}.
A symmetry of a metric space ${\cal T}=(\mathbb{Z}_N^2,\delta_N)$ is a mapping of
${\cal T}$ to itself such that the distance between points are
preserved. 
We are only interested in the set of symmetries obtained by
translations of the space $\mathbb{Z}_N^2$ (i.e., circular ``slides'' in the
vertical, horizontal directions and their combinations) because the
symmetries obtained by 
translations are exactly the ones tied to the multiplicity of
codewords in a turbo code (other possible, but not allowed, symmetries
in this paper are rotations and reflections). 
The algebraic equivalent of a translation ${\cal
  A}(k_0,k_1):\mathbb{Z}_N^2\rightarrow \mathbb{Z}_N^2$ that circularly ``slides''
to the right by  $k_0$ and upwards by $k_1$ is given by

\[
{\cal A}(k_0,k_1):(x_0,x_1)\mapsto (x_0+k_0,x_1+k_1),\quad
k_0,k_1\in \mathbb{Z}_N.
\] 

\noindent
The set of symmetry functions forms a group ${\mathcal G}$ under function
composition. Further, since the only symmetries allowed are
translations, ${\mathcal G}$ is clearly a commutative group isomorphic
to $C_N^2$ (the Cartesian product of two cyclic groups of order
$N$). An isometry of an interleaver-code $F$ is a symmetry ${\cal A}$
of ${\cal T}$ inducing ${\cal   A}:F\mapsto G$ such that
$F=G$. The set of isometries of $F$ form a subgroup ${\cal H}$ of ${\mathcal
  G}$. We say that a point $p_{x_1}\in F$ is equivalent to a
point $p_{x_2} \in F$ when there exists an isometry ${\cal
  A}$ of $F$ that maps $p_{x_1}$ to $p_{x_2}$. 

\begin{definition}
Let $F$ be an interleaver-code. The orbit of a point  $p_x\in F$ is the set
of points ${\cal   O}_{p_x}$ equivalent under the action of the
isometry group ${\cal H}$. 
\end{definition}

\begin{proposition} (Theorem 5.1~\cite{neumann}) There is just one way
 to express $F$ as the disjoint  union of a family of orbits.
\label{prop:guniq}
\end{proposition}

\begin{definition} The degree of non-linearity $\zeta(F)$ of
an interleaver $F$ is the of number of distinct orbits.
\end{definition}

\begin{proposition} All orbits have the same size.
\label{prop:osize}
\end{proposition}
\begin{proof}
This is straightforward from the fact that the only allowed symmetry functions
are translations and Proposition~\ref{prop:guniq}.
\end{proof}

\begin{definition} The degree of shift-invariance $\epsilon(F)$ of
an interleaver $F$ is the size of the orbits.
\end{definition}

\begin{proposition} The degree of non-linearity $\zeta$ and the
  degree of shift-invariance $\epsilon$ of an interleaver of length
  $N$ are related by $\zeta=N/\epsilon$. 
\label{prop:shift-nl}
\end{proposition}

\begin{proof}
This follows directly from their definitions and
Proposition~\ref{prop:osize}.
\end{proof}

Clearly $1\leq \zeta,\epsilon \leq N$. Let $l(x)=l_1x \pmod{N}$ a linear
permutation polynomial inducing an interleaver-code $L$ via $\Phi$. A possible symmetry of $L$
is $l(x)=l(x-1)-l_1$. Then the degree of non-linearity of $L$ is
$\zeta(L)=1$, i.e., the smallest possible. It is reasonable that a
randomly chosen permutation function $r(x)$ induces an
interleaver-code $R$ via  
$\Phi:r(x)\mapsto R$ whose degree of non-linearity is likely to be
$\zeta(R)=N$, i.e., the largest possible, especially if $N$ is large.   

\subsection{The degree of non-linearity of QPP interleavers}

Let us compute the degree of non-linearity of QPP interleavers
$q(x)=q_1x+q_2x^2 \pmod{N}$ by finding the isometry mappings for
the interleaver-code $Q$.

\begin{theorem}
The degree of non-linearity of a  QPP interleaver $Q$ given by
$q(x)=q_1x+q_2x^2$ is $\zeta(Q)=N/\gcd(2q_2,N)$. 
\label{th:qpp_nl_degree}
\end{theorem}

\begin{proof}
See Appendix~\ref{sec:qpp_nl_degree}.
\end{proof}

From the proof of the Theorem~\ref{th:qpp_nl_degree}, and because
$p_0=(0,0)\in Q$ if we assume a constant-free QPP, the set 

\[
{\cal O}_{p_0}=\{(k_0(i),k_1(i))| i=0,1,\ldots, \gcd(2q_2,N)
-1\}
\]
is precisely one of the orbits in $Q$. 

\begin{theorem}
The orbits for the sequence of {\em maximum-spread} interleavers in
(\ref{eq:sequence}) are interpolated by a linear curve.
\end{theorem}
\begin{proof}
This follows from the proof of Theorem~\ref{th:sequence} where the
orbits are exactly intersections between linear curves and the QPP.
\end{proof}

\begin{theorem}
Let $p_{x_1},p_{x_2}\in {{\cal O}_{(0,0)}}$ and $p_{x_1}\neq
p_{x_2}$. A lower bound on the distance $\delta_N(p_{x_1},p_{x_2})$
is $2N\gcd(2q_2,N)$.  
\label{th:orbit_sep}
\end{theorem}
\begin{proof}
Half of the distance $2N\gcd(2q_2,N)$ is from the minimum distance in set
$\{k_o(i)\}$. The other half is from the minimum distance in the set 
$\{k_1(i)\}$, which turns out to be also equally separated by
$\frac{N}{\gcd(2q_2,N)}$ by the $MCF$ Theorem in~\cite{tak:mcf}. 
\end{proof}

To find the other orbits, we only need one representative from
each. 

\begin{theorem}
A complete set of representatives for the distinct orbits of $Q$ is
\[
\{(i,f(i))|i=0,1,\ldots, N/\gcd(2q_2,N)-1\}.
\] 
\label{th:orbit_rep}
\end{theorem}

\begin{proof}
This is so because the orbits are disjoint and to have a point belong
to the same orbit, they must be no closer than $N/\gcd(2q_2,N)$ in
either coordinate by Theorem~\ref{th:orbit_sep}. These must
cover all representatives because the number of distinct orbits is
$N/\gcd(2q_2,N)$. 
\end{proof}

\begin{figure}[htbp]
  \centering
  \includegraphics[width=0.30\hsize,angle=270]{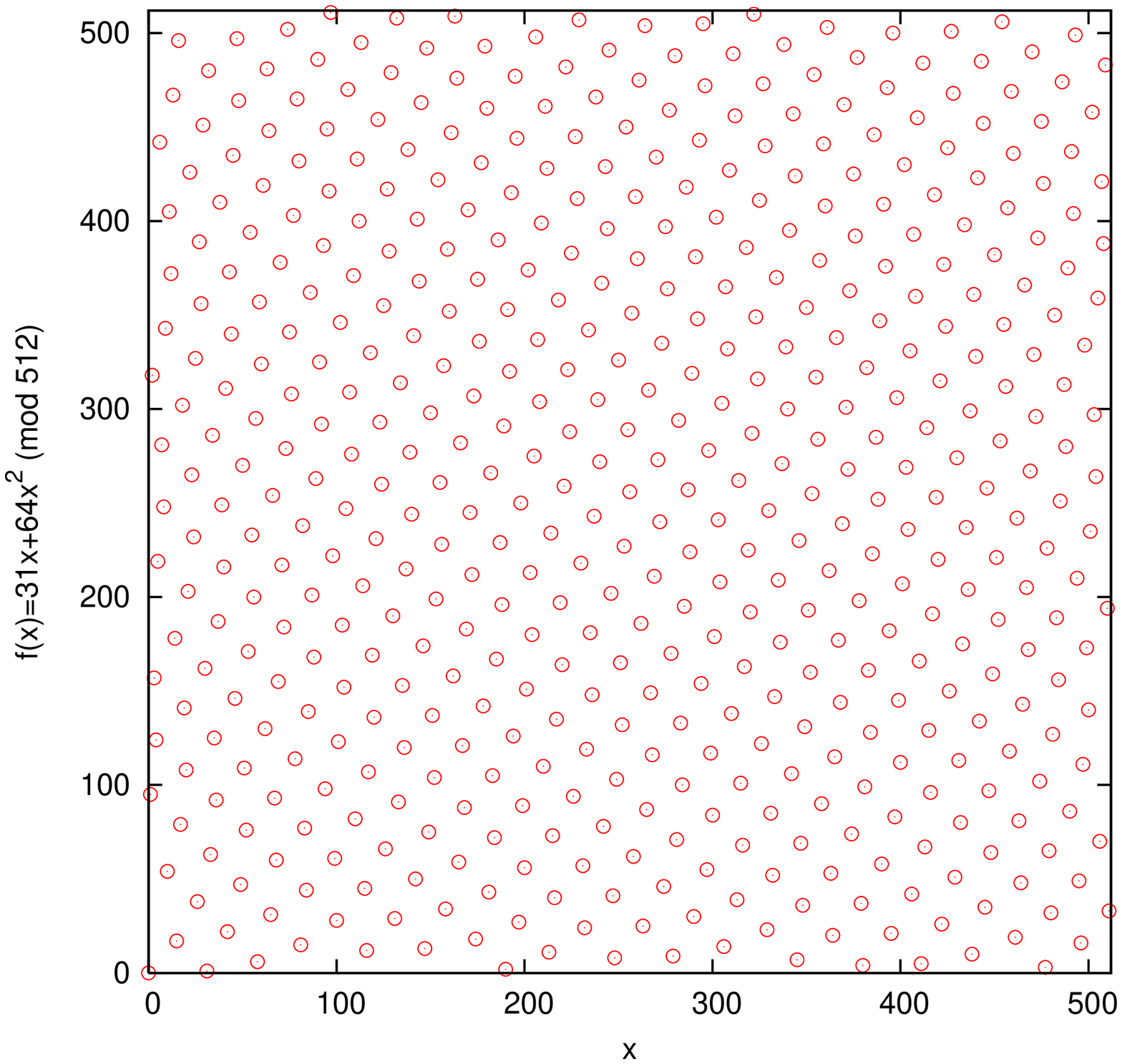}
  \includegraphics[width=0.30\hsize,angle=270]{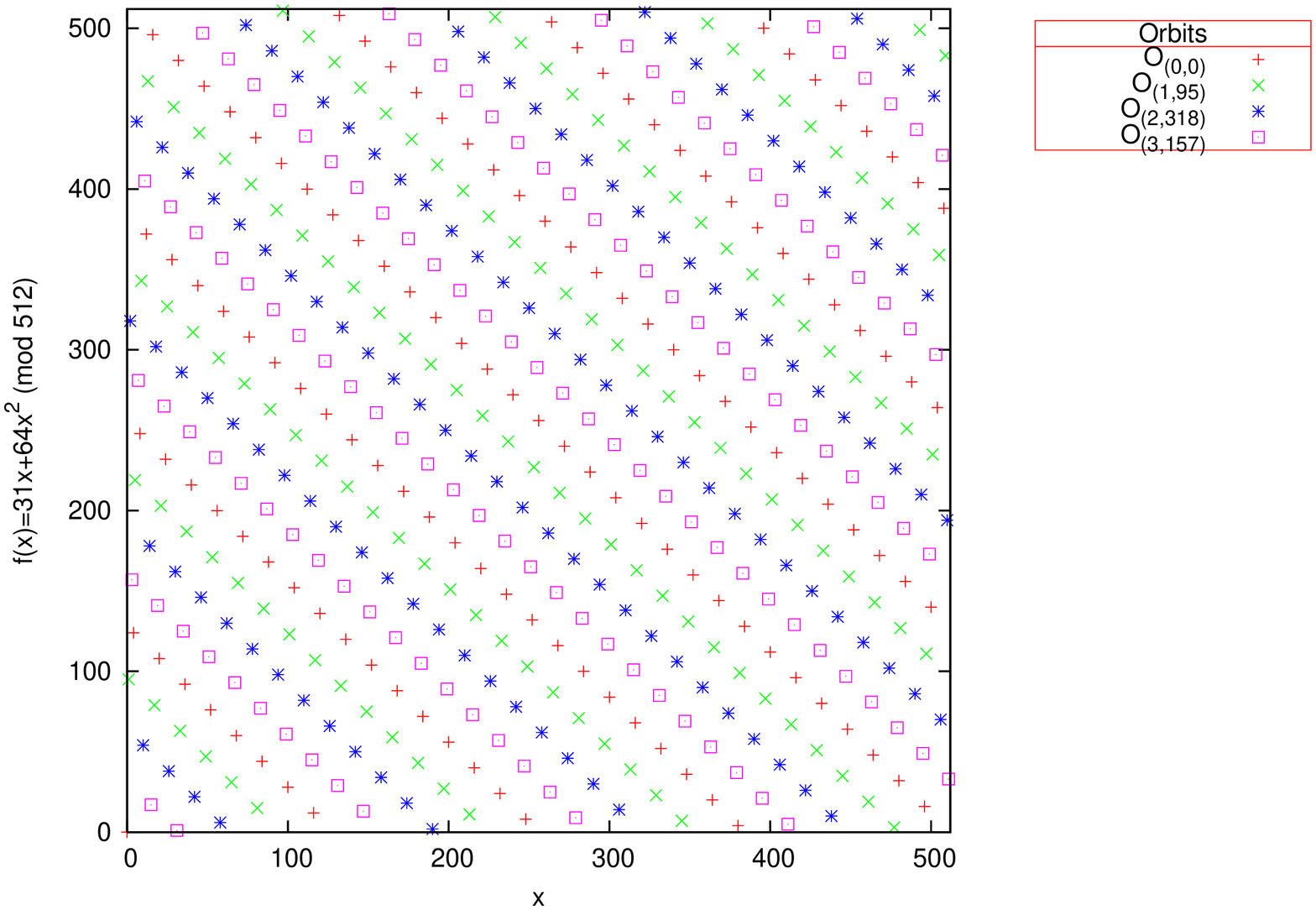}
  \caption{The interleaver $f(x)=31x+64x^2\pmod{512}$ decomposed into
    its four disjoint orbits.}
  \label{fig:MS512orbits}
\end{figure}

The decomposition of the interleaver defined by
$f(x)=31x+64x^2\pmod{512}$ into its four disjoint orbits is shown in  
Fig.~\ref{fig:MS512orbits}.  

It is interesting that the placements of the points in
both plots in Fig.~\ref{fig:MS512orbits} are exactly the same;
however, the regularity (linearity) of the interleaver gets clearly
stressed on the right plot.  

\begin{definition}Let $Q$ be a interleaver-code generated by an
  arbitrary PP. The local spread of a point $p_x\in Q$  is
\[
D_{p_x}=\min\{\delta_N(p_x,p_y)|\delta_N(p_x,p_y)\leq \sqrt{2N},\sqrt{2N} \}.
\]
\end{definition}

We have now the following Theorem on an efficient
computation procedure for the spread factor $D$ of PP interleavers.

\begin{theorem} Let $Q$ be a interleaver-code generated by an
  arbitrary PP and let $\{p_x\}$ be a set of representatives for each orbit in $Q$.
The spread factor $Q$ can be computed by
\[
D=\min\{D_{p_y}|p_y \in \{p_x\} \}.
\]
\label{th:qpp_search}
\end{theorem}
\begin{proof}
This follows from the fact that all points in a particular orbit are
equivalent under translations and a local spread cannot exceed the
upper bound on $D$.
\end{proof}

\subsection{A Refined Non-linearity Metric}

One problem with the non-linearity metric $\zeta$ is that it does not
capture the notion of orbits that are disjoint but are ``co-linear,''
i.e., there exists a linear curve interpolating them. We propose another
non-linearity metric $\zeta^\prime\leq \zeta$ that fixes part of this 
problem. A QPP $q(x)=q_1x+q_2x^2\pmod{N}$ can be decomposed
in two monomials, one corresponding to the first degree term $q_1x
\pmod{N}$ and the other to the second degree term
$q_2x^2\pmod{N}$. In most cases, we have $\gcd(q_1,N)=1$ for a valid QPP, 
which means that $q_1x \pmod{N}$ is a linear permutation polynomial
(in the case that $\gcd(q_1,N)\not =1$ a generalization is possible). A QPP can
therefore be viewed as a linear PP that is
``disturbed''\footnote{There is a similarity with DRP interleavers
  where a linear permutation polynomial is ``dithered.'' However, as
  argued in Section~\ref{sec:drp_vs_qpp}, DRP and PP interleavers are
  not the same.} by 
$q_2x^2\pmod{N}$ at every position $x$. For example, the point at
coordinate $x=0$ gets disturbed by $q_20^2=0$, the point at coordinate
$x=1$ gets disturbed by $q_21^2=q_2$ and so on. By
Theorem~\ref{th:orbit_rep}, the periodicity of the disturbance
is at most $\zeta$. The non-linearity $\zeta^\prime$ simply measures
how many distinct elements we have in the set
$\{q_2x^2\pmod{N}|x=0,1,\ldots,\zeta-1\}$, which is a very simple measure.
The values for $\zeta^\prime$ in Table~\ref{tab:msinterleavers}
reveal that they do not grow as fast as $\zeta$.  

\subsection{Low-Weight Codewords in Turbo Codes with Uniterleaved Dual
  Termination}   
\label{sec:cm}

Parallel concatenated turbo codes using trellis termination such
as the one in the 3GPP standard (both trellises are terminated but
termination bits in neither constituent codes are interleaved) are
known to suffer from low-weight codewords if a weight-one input
sequence near the end of one constituent code maps to a near-end
position in the other constituent code.  This fact is confirmed for
3GPP codes in~\cite{rosnes:tcom2005}. Similar effects for larger
input weights also happen and will be called in general as {\em edge
  effects}. The termination method will be called uninterleaved dual
termination (UDT). To minimize edge effects, the following is
desirable to be maximized: 

\[
{\cal C}=\min_{x\in \mathbb{Z}_N}\delta((N-1,N-1),(x,f(x)))
\]

We will call ${\cal C}$ the corner merit of an interleaver because
geometrically it means avoiding  points in the right upper corner of 
$F$. A lower bound on the corner merit ${\cal C}$ is guaranteed
for permutation polynomial interleavers with a constant term in the
polynomial equal to zero.

\begin{proposition}
Let the constant coefficient of the permutation polynomial
$f(x)$ be zero, then in the interleaver-code $F=\Phi(f)$, $(0,0)\in F$. This
implies ${\cal C}\geq D-2$. 
\label{prop:corner}
\end{proposition}

\begin{proof}
This follows directly from the definition of the spread factor $D$ and
the placement of one of the points in $F$ at $(0,0)$.
\end{proof}

\noindent
Therefore, a corner merit is automatically guaranteed if the spread
factor $D$ is reasonable. The concept is also valid for any
interleaver with a spread factor $D$ and one of its points placed at
$(0,0)$. For PP interleavers, the corner merit is often improved
over the lower bound in Proposition~\ref{prop:corner} by selecting a
proper constant term for the PP, i.e., searching over possible
vertical translations of the interleaver-code (it may be further
improved by examining all translations of the interleaver-code).
It is also possible that optimizing edge effects for input-weight one
sequences may degrade edge effects for other low-weight input
sequences.   

Low-weight codewords induced by UDT must have
relatively low multiplicities because they are caused by edge
effects. If the interleaver is properly designed, they have little
impact at moderate frame error rates. However, if extremely low frame
error rates are desired, a removal of edge effects looks necessary.
To remove this edge effect, there are two approaches: tail-biting
convolutional codes~\cite{weissTB} and interleaved dual
termination (IDT)~\cite{guinand_DT}.  

The tail-biting approach implies an additional complexity for encoding
and decoding. In~\cite{tak:qpp_ldpc} a construction for LDPC codes
using QPPs was proposed; it is also proved that they are
quasi-cyclic. A similar argument that is made in~\cite{tak:qpp_ldpc}
can be made to prove that turbo codes using PPs 
are quasi-cyclic when tail-biting convolutional codes are used. We
also state here without giving details that for tail-biting PP turbo
codes, the multiplicities of the low-weight codewords are typically
multiples of the degree of shift-invariance $\epsilon$.

A second form of avoiding edge-effects is by using IDT as proposed
in~\cite{guinand_DT}. Effectively, what this termination method 
produces is a sub-code of a tail-biting turbo code by choosing a
particular time in the circular code and expurgating codewords that
start/end at states other than zero. This costs a reduction in rate,
and the re-introduction of edge effects. However, because this is a sub-code
of the tail-biting form, the weight spectrum must be better than the
``mother'' tail-biting code. Further, encoding is also a little more
complex than regular turbo encoders but decoding has the same complexity.

\subsection{A Simple Metric for Permutation Polynomial Interleavers}
\label{sec:new_metric}

We introduced in~\cite{sun:tak:pp} a procedure for the selection of
QPPs matched to a choice of a constituent convolutional code that had
in mind the elimination of self-terminating input sequences of weight
$2m$ for small integer $m$'s. The procedure is, however, too
cumbersome for $m>1$. We propose now a simpler concept that has its
objective the maximization of the following metric for an 
interleaver $f(x)$: 

\begin{equation}
\Omega(f)=\ln(D(f))\zeta(f).
\end{equation}

\noindent 
In other words, we want to
maximize the product of the logarithm of the spread factor by the degree of
non-linearity of the permutation interleaver. The reasoning is that
the minimum distance of a turbo code is now known to grow at most 
logarithmically; therefore the spread factor that controls the
effective free distance should be ``rewarded'' at most
logarithmically. The non-linearity $\zeta$ is expected to have a  
proportional reduction in the multiplicities of low-weight codewords
so it is reasonable to leave it as is. 
The corner merit ${\cal C}$ is indirectly considered  because
$\Omega$ is a factor of $D$ and a lower bound for ${\cal C}$ is $D-2$
(See Proposition~\ref{prop:corner}). With this new approach, we step
aside from attempting to optimize the 
interleaver for a particular convolutional code as was done
in~\cite{sun:tak:pp}.  
This simpler measure becomes more important as we investigate more
complex polynomials of larger degrees and larger lengths. However,
this measure is still very empirical and further understanding is
desirable. As we have defined a refined non-linearity metric
$\zeta^\prime$, we have a corresponding refined metric

\begin{equation}
\Omega^\prime(f)=\ln(D(f))\zeta^\prime(f).
\end{equation}

Interleavers for turbo codes were originally constructed with random
properties with the argument that multiplicities of ``bad input
weights'' are reduced. 
We believe this is an introduction of steps to remove the notion of
a less precise notion of ``randomness'' in turbo codes to more principled
concepts. We generated the new Table~\ref{tab:maxomega} of possibly
good interleavers using this new metric and the idea of not to being
greedy in the maximization of the spread factor (see
Section~\ref{sec:Dmaximization}). The table lists polynomials for
which $\Omega^\prime$ is maximized and with a spread factor $D$ larger
or equal to $\beta ub_D(N)$; the associated interleavers will be
called $\Omega^\prime$ QPP interleavers. The threshold $\beta$ makes
sure that the spread factor does not get too small for small block lengths.  A
reasonable threshold has been determined experimentally. In fact,
as the block size increases, $\beta$ is let become smaller; this means a
maximization of the spread factor is considered less important for larger
block lengths. When multiple polynomials with the same product merit  
$\Omega^\prime$ exist, we list the one with
smallest coefficient for $f_2$ and then for $f_1$. Inverse polynomials
that are also QPP are listed when they exist. It is also listed the
constant coefficient $f_0$ that optimizes the corner merit ${\cal C}$
for $g(x)=f(x)+f_0$. The inverse polynomial $g^{-1}(x)$ is computed
by~\cite{ryu:tak:qinv} $g(x)=f^{-1}(x-f_0)$.

\begin{table}[htbp]
\centering
\caption{QPPs with best $\Omega^\prime$ and $\beta ub_D(N)\leq D$.}
\begin{tabular}{|r|r|c|c|r|r|c|c|} \hline
$N$ & $f_0$ & $f(x)$ & $f^{-1}(x)$ & $D$ &$\Omega^\prime$ & $\zeta^\prime$ & $\beta$\\ \hline
40  & 6  & $x+10x^2$   & $21x+10x^2$    &4  & 2.77 & 2 & 0.45\\
80  & 72 & $9x+20x^2$  & $49x+20x^2$    &10 & 4.61 & 2 & 0.45\\
128 & 89 & $7x+16x^2$  & $55x+16x^2$    &8  & 6.24 & 3 & 0.45\\
160 & 115& $9x+20x^2$  & $9x+60x^2$     &10 & 6.91 & 3 & 0.45\\
256 & 240& $15x+32x^2$ & $239x+32x^2$   &16 & 8.32 & 3 & 0.45\\
320 & 304& $19x+40x^2$ & $59x+40x^2$    &20 & 8.99 & 3  & 0.45\\
400 & 375& $7x+40x^2$  & $343x+120x^2$  &16 & 13.86& 5  & 0.45\\
408 & 273& $25x+102x^2$& $253x+102x^2$  &24 & 6.36 & 2  & 0.45\\
512 & 433& $15x+32x^2$ & $239x+32x^2$   &16 & 11.09& 4 & 0.45\\
640 & 549& $19x+ 40x^2$& $219x+ 200x^2$ &20 & 11.98& 4 & 0.45\\
752 & 619& $23x+ 94x^2$& $327x+94x^2$   &26 & 9.77& 3  & 0.45\\
800 & 786& $17x+ 80x^2$& $753x+ 240x^2$ &32 & 17.33& 5 & 0.45\\ 
1024& 992& $31x+64x^2$ & $991x+64x^2$   &32 & 13.86& 4  & 0.45\\
1280&1248& $39x+80x^2$ & $279x+80x^2$    &40 & 14.76& 4 & 0.45\\
1504&1463& $23x+94x^2$ & --  &26& 13.03 & 4  & 0.45\\
1600&1169& $17x+80x^2$ & $753x+240x^2$  & 32& 20.79 & 6  & 0.45\\
2048&1315& $31x+ 64x^2$& $991x+ 64x^2$  &32 & 24.26 & 7 & 0.30\\
2560&2377& $39x+80x^2$ & $2199x+720x^2$ &40 & 25.82 & 7 & 0.30\\
3200&2328& $17x+80x^2$&  $2353x+240x^2$  &32 & 31.19 & 9    & 0.30\\
4096&1332& $31x+64x^2$& $991x+64x^2$ &32 & 41.59 & 12 & 0.30\\
5472&3104& $77x+114x^2$&  --            &36 & 28.67 & 8 & 0.30\\
8192&1084& $31x+64x^2$& $3039x+2112x^2$ &32 & 79.71 & 23& 0.30\\
\hline
\end{tabular}
\label{tab:maxomega}
\end{table}

\subsection{Permutation Polynomials of Degrees Larger than Two}

We have mentioned that while there are linear {\em maximum-spread}
interleavers for every $N=2n^2$, this is not true for quadratic
permutation polynomials.  For example, if $N=200=2\times 10^2$ there
are no {\em maximum-spread} interleavers of second degree. However,
the cubic permutation 
polynomial $f(x)=59x+60x^2+20x^3\pmod{200}$ generates a {\em
  maximum-spread} interleaver. Further, interleaver lengths not
admitting quadratic permutation polynomials at all may have permutation
polynomials of larger degrees. For example, if $N=5$, there are no
quadratic permutation polynomials but $c(x)=x^3\pmod{5}$ is a cubic
permutation polynomial of irreducible degree. This means maybe it is
worth investigating polynomials of larger degrees because the
complexity of evaluating the polynomials only grows linearly with the
degree using Horner's rule~\cite[page 93]{Gathen}. The cubic
polynomial $f(x)=59x+60x^2+20x^3\pmod{200}$ can be evaluated at every
point $x$ with three multiplications and two additions using
$f(x)=(59+(60+20x)x)x \pmod{200}$. 
Further, it would be interesting if an efficient implementation that
sequentially generates all evaluations $f(0),f(1),f(2),\ldots$ that
requires only additions and comparisons for arbitrary polynomial
degrees, generalizing the idea in~\cite{jpl2}, is possible. To determine the
coefficients for a permutation polynomial of an arbitrary degree and
an arbitrary $N$, there is an easy sufficient condition by using
Theorem~2.3 and Corollary~2.5 in~\cite{sun:tak:pp}. 

\section{Numerical Results}
\label{sec:nr}

\subsection{Distance Spectra of some Example Codes}
\label{sec:ds}

Algorithms for computing the true distance spectra of turbo codes
such as in~\cite{gar:jsac2001} and~\cite{rosnes:tcom2005} are very useful to help
analysis. We used the algorithm in~\cite{gar:jsac2001} to compute the first 20
smallest distances and respective multiplicities for a few turbo codes
using interleavers in Tables~\ref{tab:ms128}-\ref{tab:ms512}. We
constrained the search for input weights at most 10. The turbo codes
are parallel concatenated codes of nominal rate 1/3. We used two types
of constituent codes: the 8-state  constituent convolutional code with 
generator matrix $[1\quad (1+D+D^3)/(1+D^2+D^3)]$, also denoted in
octal notation as (13,15) and the 16-state constituent convolutional
code with generator matrix  $[1\quad (1+D+D^2+D^4)/(1+D^3+D^4)]$, also
denoted in octal notation as (23,35). The trellis termination method
was the same as for the 3rd Generation Partnership Project (3GPP)
standard~\cite{3gpp}, i.e., UDT. 

\begin{table}[htbp]
\centering
\caption{Length $N=128$}
\begin{tabular}{ |c |c |c |c |c |c |c |c |c |c |c |c |c |c |c |c |c |c
    |c |c |c |}\hline
\multicolumn{21}{|c|}{  8-state  }  \\ \hline
\multicolumn{21}{|c|}{MS QPP $f(x)=15x+32x^2\pmod{N}$  
  $\epsilon=64$}  \\ \hline
$i$ & {\bf 0} & 1 & 2 & 3 & {\bf 4} & 5 & 6 & 7 & 8 & 9 & 10 & 11 & 12 & 13 & 14 & 15 & 16 & 17 & 18 & 19 \\ \hline\hline
$d_i$ & {\bf 16} & 18 & 19 & 20 & {\bf 21} & 22 & 23 & 24 & 25 & 26 & 27 & 28 & 29 & 30 & 31 & 32 & 33 & 34 & 35 & 36 \\ \hline
$N_i$ & {\bf 1} & 1 & 1 & 2 & {\bf 55} & 2 & 7 & 8 & 9 & 16 & 27 & 120
& 53 & 359 & 526 & 391 & 775 & 1368 & 1865 & 2542 \\ \hline \hline

\multicolumn{21}{|c|}{$\Omega^\prime$ QPP $f(x)=7x+16x^2\pmod{N}$ 
  $\epsilon=32$}  \\ \hline
$i$ &{\bf 0} & 1 & 2 & 3 & 4 & 5 & 6 & 7 & {\bf 8} & 9 & 10 & 11 & 12 & 13 & 14 & 15 & 16 & 17 & 18 & 19 \\ \hline\hline
$d_i$ & {\bf 14} & 15 & 16 & 18 & 19 & 20 & 21 & 22 & {\bf 23} & 24 & 25 & 26 & 27 & 28 & 29 & 30 & 31 & 32 & 33 & 34 \\ \hline
$N_i$ & {\bf 2} & 1 & 1 & 2 & 4 & 3 & 8 & 8 & {\bf 33} & 37 & 24 & 123 & 72 & 198
& 111 & 296 & 578 & 731 & 1240 & 1822 \\ \hline

\multicolumn{21}{|c|}{$\Omega^\prime$ QPP $f(x)=89+7x+16x^2\pmod{N}$ 
  $\epsilon=32$}  \\ \hline
$i$ & {\bf 0} & 1 & 2 & 3 & 4 & 5 & 6 & {\bf 7} & 8 & 9 & 10 & 11 & 12 & 13 & 14 & 15 & 16 & 17 & 18 & 19 \\ \hline\hline
$d_i$ & {\bf 12} & 16 & 18 & 19 & 20 & 21 & 22 & {\bf 23} & 24 & 25 & 26 & 27 & 28 & 29 & 30 & 31 & 32 & 33 & 34 & 35 \\ \hline
$N_i$ & {\bf 1} & 1 & 3 & 2 & 1 & 7 & 8 & {\bf 34} & 32 & 15 & 117 & 61 & 184 &
119 & 289 & 532 & 651 & 1166 & 1731 & 3023 \\ \hline\hline

\multicolumn{21}{|c|}{  16-state  }  \\ \hline
\multicolumn{21}{|c|}{MS QPP $f(x)=15x+32x^2\pmod{N}$ 
  $\epsilon=64$ }  \\ \hline
$i$ &    {\bf 0} & 1 & 2 & 3 & 4 & 5 & 6 & 7 & 8 & 9 & 10 & 11 & {\bf 12} & 13 & 14 & 15 & 16 & 17 & 18 & 19 \\ \hline\hline
$d_i$ & {\bf 17} & 18 & 19 & 20 & 21 & 22 & 23 & 24 & 25 & 26 & 27 &
28 & {\bf 29} & 30 & 31 & 32 & 33 & 34 & 35 & 36 \\ \hline
$N_i$ & {\bf 1} & 3 & 2 & 3 & 1 & 1 & 3 & 1 & 6 & 9 & 14 & 13 & {\bf
  52} & 28 & 199 & 78 & 79 & 154 & 235 & 422 \\ \hline

\multicolumn{21}{|c|}{ $\Omega^\prime$ QPP $f(x)=7x+16x^2\pmod{N}$ 
  $\epsilon=32$}  \\ \hline
$i$ & {\bf 0} & 1 & 2 & 3 & 4 & 5 & {\bf 6} & 7 &  8 & 9 & 10 & 11 & 12 & 13 & 14 & 15 & 16 & 17 & 18 & 19 \\ \hline\hline
$d_i$ & {\bf 17} & 21 & 23 & 24 & 25 & 26 & {\bf 27} & 28 &  29 & 30 & 31 & 32 & 33 & 34 & 35 & 36 & 37 & 38 & 39 & 40 \\ \hline
$N_i$ & {\bf 1} & 2 & 2 & 2 & 4 & 5 & {\bf 31} & 18 &  55 & 23 & 97 & 103 & 133 & 253 & 234 & 596 & 722 & 1261 & 1923 & 3119 \\
\hline

\multicolumn{21}{|c|}{$\Omega^\prime$ QPP $f(x)=89+7x+16x^2\pmod{N}$ 
  $\epsilon=32$} \\ \hline
$i$ & {\bf 0} & 1 & 2 & 3 & {\bf 4} & 5 & 6 & 7 & 8 & 9 & 10 & 11 & 12 & 13 & 14 & 15 & 16 & 17 & 18 & 19 \\ \hline\hline
$d_i$ & {\bf 23} & 24 & 25 & 26 & {\bf 27} & 28 & 29 & 30 & 31 & 32 & 33 & 34 & 35 & 36 & 37 & 38 & 39 & 40 & 41 & 42 \\ \hline
$N_i$ & {\bf 3} & 1 & 4 & 5 & {\bf 31} & 9 & 56 & 16 & 93 & 101 & 109 & 233 & 190 & 577 & 655 & 1099 & 1731 & 2920 & 4407 & 6408 \\

\hline

\end{tabular}
\label{tab:ms128}
\end{table}

\begin{table}[htbp]
\centering
\caption{Length $N=512$}
\begin{tabular}{ |c |c |c |c |c |c |c |c |c |c |c |c |c |c |c |c |c |c
    |c |c |c |}\hline
\multicolumn{21}{|c|}{8-state}  \\ \hline
\multicolumn{21}{|c|}{MS QPP $f(x)=31x+64x^2\pmod{N}$  $\epsilon=128$}  \\ \hline
$i$ & {\bf 0} & 1 & 2 & 3 & 4 & 5 & {\bf 6} & 7 & 8 & 9 & 10 & 11 & 12 & 13 & 14 & 15 & 16 & 17 & 18 & 19 \\ \hline\hline
$d_i$ & {\bf 27} & 28 & 29 & 30 & 31 & 32 & {\bf 33} & 34 & 35 & 36 & 37 & 38 & 39 & 40 & 41 & 42 & 43 & 44 & 45 & 46 \\ \hline
$N_i$ & {\bf 1} & 2 & 5 & 1 & 8 & 3 & {\bf 126} & 4 & 14 & 14 & 17 & 960 & 1923 & 41 & 304 & 99 & 730 & 1113 & 539 & 406 \\
\hline \hline

\multicolumn{21}{|c|}{$\Omega^\prime$ QPP $f(x)=15x+32x^2\pmod{N}$  $\epsilon=64$}  \\ \hline
$i$ & {\bf 0} & 1 & 2 & 3 & 4 & 5 & 6 & 7 & 8 & 9 & 10 & 11 & 12 & 13 & 14 & 15 & {\bf 16} & 17 & 18 & 19 \\ \hline\hline
$d_i$ & {\bf 16} & 21 & 22 & 24 & 25 & 26 & 27 & 28 & 29 & 30 & 31 & 32 & 33 & 34 & 35 & 36 & {\bf 37} & 38 & 39 & 40 \\ \hline
$N_i$ & {\bf 1} & 1 & 1 & 1 & 4 & 2 & 2 & 3 & 5 & 7 & 11 & 13 & 18 & 16 & 36 & 30 & {\bf 114} & 200 & 665 & 543 \\
\hline

\multicolumn{21}{|c|}{$\Omega^\prime$ QPP $f(x)=433+15x+32x^2\pmod{N}$  $\epsilon=64$}  \\ \hline
$i$ & {\bf 0} & 1 & 2 & 3 & 4 & 5 & 6 & 7 & 8 & 9 & 10 & 11 & 12 & 13 & 14 & {\bf 15} & 16 & 17 & 18 & 19 \\ \hline\hline
$d_i$ & {\bf 20} & 21 & 24 & 25 & 26 & 27 & 28 & 29 & 30 & 31 & 32 & 33 & 34 & 35 & 36 & {\bf 37} & 38 & 39 & 40 & 41 \\ \hline
$N_i$ & {\bf 1} & 1 & 1 & 5 & 2 & 1 & 2 & 5 & 9 & 15 & 9 & 13 & 22 & 27 & 34 & {\bf 114} & 196 & 669 & 530 & 430 \\
\hline

\multicolumn{21}{|c|}{Degree-6 PP 512}  \\ \hline
$i$ & {\bf 0} & 1 & 2 & 3 & 4 & 5 & 6 & 7 & 8 & {\bf 9} & 10 & 11 & 12 & 13 & 14 & 15 & 16 & 17 & 18 & 19 \\ \hline\hline
$d_i$ & {\bf 27} & 28 & 29 & 30 & 31 & 32 & 33 & 34 & 35 & {\bf 36} & 37 & 38 & 39 & 40 & 41 & 42 & 43 & 44 & 45 & 46 \\ \hline
$N_i$ & {\bf 2} & 2 & 5 & 1 & 5 & 2 & 9 & 3 & 14 & {\bf 246} & 79 &
505 & 563 & 58 & 76 & 371 & 617 & 432 & 591 & 1324 \\ \hline \hline

\multicolumn{21}{|c|}{16-state}\\ \hline
\multicolumn{21}{|c|}{MS QPP $f(x)=31x+64x^2\pmod{N}$ 
  $\epsilon=128$}  \\ \hline
$i$ & {\bf 0} & 1 & 2 & 3 & 4 & 5 & 6 & 7 & 8 & 9 & 10 & 11 & 12 & 13
& 14 & 15 & 16 & 17 & {\bf 18} & 19 \\ \hline\hline
$d_i$ & {\bf  26} & 27 & 29 & 31 & 33 & 34 & 35 & 36 & 37 & 38 & 39 &
40 & 41 & 42 & 43 & 44 & 45 & 46 & {\bf 47} & 48 \\ \hline
$N_i$ & {\bf 2}& 1 & 1 & 1 & 3 & 1 & 4 & 3 & 3 & 4 & 2 & 7 & 14 & 9 &
18 & 25 & 28 & 41 & {\bf 151} & 184 \\ \hline

\multicolumn{21}{|c|}{$\Omega^\prime$ QPP $f(x)=15x+32x^2\pmod{N}$ 
  $\epsilon=64$}  \\ \hline
$i$ & {\bf 0} & 1 & 2 & 3 & 4 & 5 & 6 & 7 & 8 & 9 & 10 & 11 & 12 & 13 & 14 & 15 & 16 & 17 & 18 & 19 \\ \hline\hline
$d_i$ & {\bf 17} & 19 & 23 & 25 & 26 & 27 & 28 & 29 & 30 & 31 & 32 & 33 & 34 & 35 & 36 & 37 & 38 & 39 & 40 & 41 \\ \hline
$N_i$ & {\bf 1} & 1 & 1 & 1 & 1 & 1 & 1 & 1 & 1 & 4 & 1 & 3 & 3 & 2 & 12 & 5 & 14 & 16 & 13 & 17 \\
\hline

\multicolumn{21}{|c|}{$\Omega^\prime$ QPP $f(x)=433+15x+32x^2\pmod{N}$ 
  $\epsilon=64$} 
 \\ \hline
$i$ & {\bf 0} & 1 & 2 & 3 & 4 & 5 & 6 & 7 & 8 & 9 & 10 & 11 & 12 & 13 & 14 & 15 & 16 & 17 & 18 & 19 \\ \hline\hline
$d_i$ & {\bf 18} & 19 & 23 & 25 & 26 & 27 & 28 & 29 & 31 & 32 & 33 & 34 & 35 & 36 & 37 & 38 & 39 & 40 & 41 & 42 \\ \hline
$N_i$ & {\bf 1} & 1 & 1 & 1 & 1 & 2 & 1 & 1 & 4 & 1 & 6 & 2 & 3 & 7 & 6 & 14 & 19 & 11 & 16 & 16 \\
\hline

\multicolumn{21}{|c|}{Degree-6 PP 512}  \\ \hline
$i$ & {\bf 0} & 1 & 2 & 3 & 4 & 5 & {\bf 6} & 7 & 8 & 9 & 10 & 11 & 12 & 13 & 14 & 15 & 16 & 17 & 18 & 19 \\ \hline\hline
$d_i$ & {\bf 27} & 28 & 29 & 30 & 31 & 33 & {\bf 35} & 36 & 37 & 38 & 39 & 40 & 41 & 42 & 43 & 44 & 45 & 46 & 47 & 48 \\ \hline
$N_i$ & {\bf 2} & 1 & 2 & 1 & 1 & 3 & {\bf 63} & 1 & 3 & 6 & 3 & 1 & 6 & 12 & 10 & 19 & 21 & 33 & 45 & 48 \\
\hline

\end{tabular}
\label{tab:ms512}
\end{table}

We make the following observations:
\begin{itemize}
\item Improving the corner merit ${\cal C}$ has mixed effects. It
  improves the minimum distance (highlighted in the tables) for
  certain cases but it degrades for others as commented in
  Section~\ref{sec:cm}.
\item The first spectral line with a high multiplicity has a
  multiplicity very close to the degree of shift invariance
  $\epsilon$ as predicted in Section~\ref{sec:cm}. 
\item The $\Omega^\prime$ interleavers appear in general to improve
  the first spectral line that has a high multiplicity. However, the
  comparison is not straightforward because the distance spectra is
  truncated to the first 20 lines.
\end{itemize}

\subsection{Computer Simulation Results}
\label{sec:simulations}

The error performances of turbo codes using the {\em maximum-spread}
QPP interleavers in Table~\ref{tab:msinterleavers} as well as
interleavers in Tables~\ref{tab:maxqpp} and~\ref{tab:maxomega} are
examined. The structure of the simulated turbo codes is the same as
in the previous section. However, we simulated for both UDT and
IDT. The true code-rates for 8-state constituent codes are $N/(3N+12)$
and $(N-6)/3N$ for UDT and IDT, respectively. The true code-rates for
16-state constituent codes are $N/(3N+16)$ and $(N-8)/3N$ for UDT and
IDT, respectively. 

The true code-rate is used to 
compute the signal-to-noise ratio $(E_b/N_0)$. We used BPSK modulation and assumed
an additive white Gaussian noise (AWGN) channel. The decoding was performed
with log-MAP decoding and simulated until at least 50 frame
errors had been counted, unless otherwise noted. 

\subsubsection{Comparison with DRP Interleavers}
\label{sec:drp_vs_qpp}

\begin{figure}[htbp]
  \centering
  \includegraphics[width=0.6\hsize]{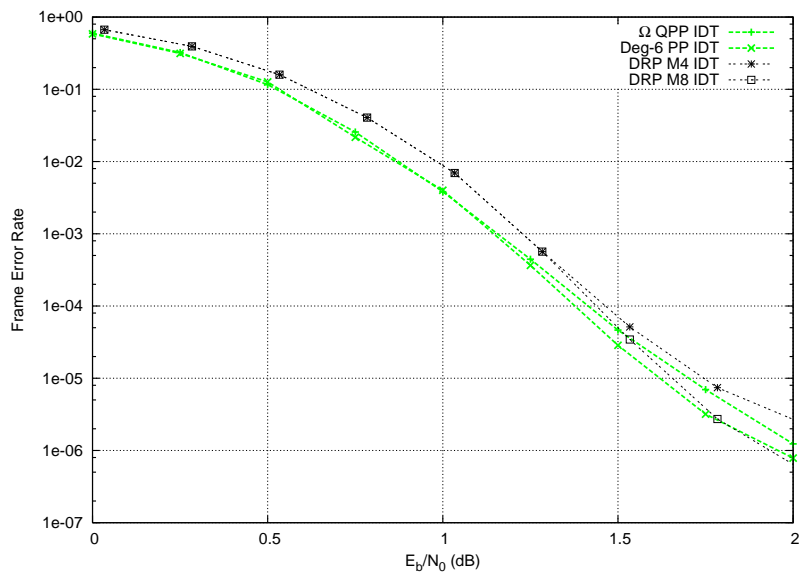}
  \caption{Length $N=512$ PP interleavers compared with DRP
    interleavers in turbo codes with 8-state constituent convolutional
    codes and 16 decoding iterations.} 
  \label{fig:drpvsqpp}
\end{figure}

We compare first length $N=512$  PP interleavers under similar
conditions with DRP interleavers in~\cite{crozier_DRP}. 
The iterative decoding method used in~\cite{crozier_DRP} is simpler than
log-MAP but worse by about 0.1dB at lower SNRs. The DRP curves were
obtained directly from~\cite{crozier_DRP} but adjusted in SNR to the
true code-rate. We used 16 log-MAP decoding iterations to match their
number of iterations. The simulation curves are compared in
Fig.~\ref{fig:drpvsqpp}.

\noindent
There is a close match in performance with the DRP interleaver with
parameter $M=4$ at higher SNRs. The similarity in performance is not
quite surprising if we consider that both PP and DRP constructions are
based on number congruences. A deeper study of their resemblance is
left for a future work, however, we provide a few insights of why QPP
interleavers are expected to be close to DRP interleavers with $M=4$
for this interleaver length of $N=512$.

The parameter $M$ in a DRP interleaver is the length of an arbitrary
permutation of $M$ letters. Explaining only the essence, a DRP
interleaver can be understood as a modification of a linear
interleaver by using the permutation of $M$ letters. We
show that PP and DRP constructions are in general different, except
when $M=1$. When $M=1$, DRP interleavers are what we call linear
interleavers, i.e., PPs of first degree because no modification of a
linear interleaver is possible if $M=1$. As $M$ grows, DRP 
interleavers can become arbitrarily ``random'' in the sense 
that when $M=N$, arbitrary interleaving functions are possible. The
class of PP interleavers cannot generate arbitrary interleaving functions
regardless of the degree of the permutation polynomial. This is
easily proven by counting arguments but also because {\em all}
interleavers generated by PPs are maximum
contention-free~\cite{tak:mcf} but an arbitrary interleaving function
is generally not (maximum) contention-free~\cite{nim:isit04}.   

Factoring out the construction algorithm, which is similar in
complexity and nature, one may characterize PP and DRP interleavers by
their input parameters (two vectors of length $M$ for DRP interleavers
and the coefficients of the polynomials for PP interleavers). We
propose to measure the ``randomness'' of DRP interleavers by computing
the entropy $E$ of the two input permutations of $M$ letters. Considering
that an arbitrary sequence of $M$ letters can be chosen, then the
entropy is $2M\log_2(M)$ bits, which is independent of the interleaver
length. For an interleaver length of $N$, the entropy of the
coefficients in $q(x)=q_1x+q_2x^2 \pmod{N}$ is $\deg(q(x))\log_2(N/2)$
bits assuming that $N$ is a power of 2 and $q_i$'s are either even or
odd~\cite{sun:tak:pp}. Now observing the previous matching in
Fig.~\ref{fig:drpvsqpp} for DRP $(M=4)$ and QPP, we have
$E(DRP)=8\log_2(4)=16$ bits and $E(QPP)=2\log_2(256)=16$ bits 
with a perfect match.\footnote{It may be argued that the input
  complexity of DRP 
  interleavers is $\log_2(N)$ higher because it requires the
  specification of a base linear interleaver that includes a constant
  coefficient.} Would this be a coincidence? The DRP interleaver
with $M=8$ has an entropy of $16\log_2(8)=48$ bits, which would
correspond to a degree-6 PP. We obtained a very close
error performance with a degree six permutation polynomial. 

\begin{equation}
f(x)=15x+ 16x^2+ 128x^3+  32x^4+ 32x^5+ 64x^6 \pmod{512}
\label{eq:sixth}
\end{equation}

\noindent
The polynomial in (\ref{eq:sixth}) has a degree of non-linearity
$\zeta=8$, a refined degree of non-linearity $\zeta^\prime=6$ and the
orbit size is $\epsilon=64$. Its spread factor is $D=26$ and has a refined
product merit of $\Omega^\prime=19.55$. The corresponding performance
curve is shown in Fig.~\ref{fig:drpvsqpp}. We used interleaved dual
termination (IDT)~\cite{guinand_DT} for this code, which gave a small
advantage over  UDT at high SNRs (it is not shown in the plot but
about a factor of 2 at 2.0dB). 

In summary:

\begin{itemize}
\item The maximum-spread interleaver ($f(x)=31x+64x^2$) is mostly indifferent
  to the termination methods and very similar in performance to the
  DRP $M=4$ and IDT.
\item The $\Omega^\prime$ interleaver ($f(x)=15x+32x^2$) has a significant
  improvement with IDT over UDT and is a little better than the maximum-spread code. 
\item The degree-6 interleaver with IDT and DRP $M=8$ interleaver have
  mostly the same performance. 
\end{itemize}

\subsubsection{Performance Curves for 8-State Constituent Codes}

We plot next the FER  curves for the first four {\em maximum-spread}
interleavers in Table~\ref{tab:msinterleavers} using UDT and the
corresponding constant-free codes in Table~\ref{tab:maxomega} using
IDT for turbo 
codes using the 8-state (13,15) convolutional codes and eight decoding 
iterations. Similarly as in ~\cite{tak:mcf}, excellent error
performance is obtained down to at least an FER of $10^{-4}$. There is
noticeable improvement in FER for  codes using $\Omega^\prime$
interleavers as the block length increases. Also, a truncated union bound is
plotted using the 20 first spectral lines for two of the codes using
maximum-spread interleavers (MS QPP $N=128$ and MS QPP $N=512$).

\begin{figure}[htbp]
  \centering
  \includegraphics[width=0.6\hsize]{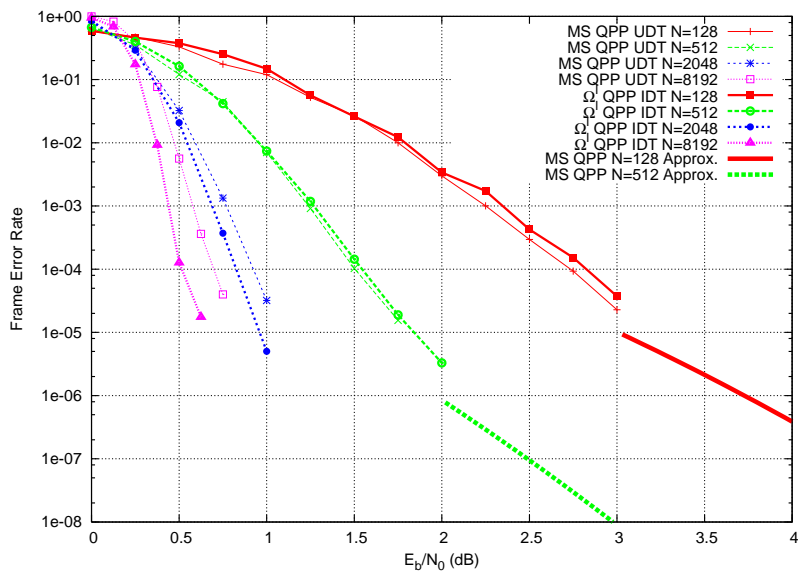}
  \caption{FER curves for turbo codes with maximum-spread
    QPP interleavers, 8-state convolutional codes, and eight decoding
    iterations.} 
  \label{fig:MSQPP_8st_8it}
\end{figure}

The refined product merit effectively avoids the selection of bad
interleavers. For example, the permutation polynomial

\[
31x+64x^2+64x^3+32x^4+64x^5+32x^6 \pmod{512}
\]

\noindent
has a spread factor $D=32$ (a maximum-spread interleaver), a degree of
non-linearity $\zeta=8$, a simple product measure of $\Omega=27.73$
but a refined degree of non-linearity $\zeta^\prime=2$, and a refined
product measure of $\Omega^\prime=6.93$. The simulation 
performance at 1.75dB (not shown in plots) gives an FER of $6\times
10^{-5}$, clearly inferior to the maximum-spread interleaver of second
degree that has a larger $\zeta^\prime=3$.

\subsubsection{Performance Curves for 16-State Constituent Codes}

An impressive error performance~\cite{crozier_MPEG} with FERs around
$3\times 10^{-8}$ at an SNR of 1.25dB was demonstrated for a DRP
interleaver with $M=8$ in a turbo code using
the (23,35) 16-state constituent code, IDT, and 16 decoding
iterations. We show that a similar result is obtained with the QPP interleaver
$f(x)=1463+23x+94x^2\pmod{1504}$ in Table~\ref{tab:maxomega} with the
same constituent code, UDT, and 16 log-MAP decoding iterations. We
have not simulated the constant-free case with IDT but it should
provide an even better error performance. Other codes with
interleavers in Tables~\ref{tab:maxqpp} (MS) and
Tables~\ref{tab:maxomega} ($\Omega^\prime$) were also simulated. 

A genie stopper for the decoding iterations (the iterations are stopped when
the decoded information sequence completely agrees with the
transmitted one) was used when simulating points at very low FERs. For
the length $N=8192$ $\Omega^\prime$ QPP code, at 0.5dB, close to 6 
million frames were simulated with no frame errors with a regular turbo
decoder; additionally, almost 40 million frames were simulated with a
genie stopper for the iterations, resulting in only three frame errors. 
Similarly, other points at very low FERs were simulated with the
decoding parameters in Table~\ref{tab:sim}.

\begin{table}[htbp]
\centering
\caption{Simulation Parameters for low FERs in Fig.~\ref{fig:MSQPP_16st_16it}}
\begin{tabular}{|l|r|r|} \hline
\multicolumn{3}{|c|}{$N=512\quad$ MS QPP IDT 2.00dB}\\ \hline
        & Frames      & Frame Errors \\ \hline
Regular & 74,151,303  &  1     \\
Genie   & 657,736,347 &  7        \\
Total   & 731,887,650 &  8     \\  \hline
\multicolumn{3}{|c|}{$N=512\quad$ MS QPP UDT 2.00dB}\\ \hline
        & Frames    & Frame Errors \\ \hline
Regular & 56,273,401  &  10     \\
Genie   & 0           &  0        \\
Total   & 56,273,401  &  10     \\  \hline

\multicolumn{3}{|c|}{$N=512\quad$ $\Omega^\prime$ QPP IDT 2.00dB}\\ \hline
        & Frames    & Frame Errors \\ \hline
Regular & 195,667,074 &  11     \\
Genie   & 0         &  0        \\
Total   & 195,667,074 &  11     \\  \hline
\multicolumn{3}{|c|}{$N=1504\quad$ $\Omega^\prime$ QPP UDT 1.25dB}\\ \hline
        & Frames   & Frame Errors \\ \hline
Regular & 25,603,477 &  1          \\
Genie   & 180,873,414&  3          \\
Total   & 206,476,891 & 4          \\  \hline
\multicolumn{3}{|c|}{$N=8192\quad$ $\Omega^\prime$ QPP IDT 0.50dB}\\ \hline
        & Frames    & Frame Errors\\ \hline
Regular & 5,887,701   & 0           \\
Genie   & 39,323,725  & 3           \\
Total   & 45,211,426  & 3           \\ \hline
\multicolumn{3}{|c|}{$N=8192\quad$ MS QPP IDT 0.50dB}\\ \hline
        & Frames    & Frame Errors\\ \hline
Regular & 4,299,436   & 0           \\
Genie   & 29,305,054  & 3           \\
Total   & 33,604,490  & 3           \\ \hline
\end{tabular}
\label{tab:sim}
\end{table}

\begin{figure}[htbp]
  \centering
  \includegraphics[width=0.6\hsize]{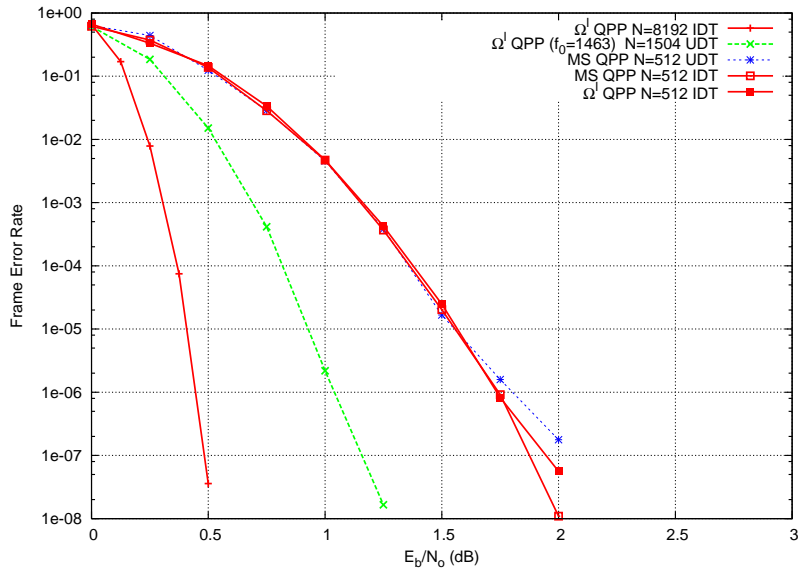}
  \caption{FER curves for turbo codes with QPP interleavers, 16-state
    constituent convolutional codes, and 16 decoding iterations.}
  \label{fig:MSQPP_16st_16it}
\end{figure}

Although there was an evident improvent by using $\Omega^\prime$
polynomials for larger block lengths using 8-state constituent codes, 
there is not a clear difference for the length $N=8192$ using 16-state
constituent codes within the reach of simulations. 

In addition, we observe the following for length $N=512$:

\begin{itemize}
\item Codes using both the maximum-spread interleaver
  ($f(x)=31x+64x^2$) and the $\Omega^\prime$ ($f(x)=15x+32x^2$)
  interleaver benefit significantly with IDT.
\item  The code with the maximum-spread interleaver and IDT 
is better than the code with $\Omega^\prime$ ($f(x)=15x+32x^2$). 
\item  The code with the maximum-spread interleaver and IDT has a
  performance very close to the best interleaver in~\cite{crozier2}, a 
  dithered diagonal interleaver with identical spread factor of
  $D=32$. 
\end{itemize}

In summary, for resource-constrained applications, the combination of
8-state constituent codes, a maximum of 8 decoding iterations and QPP
interleavers give excellent performance down to FERs around $10^{-4}$
or smaller from short to medium information block sizes of up to 8192
bits. If 16-state constituent codes and a maximum of 16 decoding
iterations are acceptable then QPP interleavers give impressive
performance down to FERs close to $10^{-8}$ with little or no signs
of error-floors.  

\section{Conclusions}
\label{sec:conclusions}

The recently proposed construction of PP
interleavers~\cite{sun:tak:pp} and the subclass of QPP interleavers
yield very good error rate performance for turbo codes in a number of
practical examples~\cite{tak:mcf,jpl2}. Interesting properties of QPP
interleavers have been studied
earlier~\cite{sun:tak:pp,ryu:tak:qinv,tak:mcf}; however, their 
characterization is still in its infancy with many open questions. In
this paper, the spread factor of QPP interleavers was studied. An
infinite sequence of QPP 
interleavers achieving the {\em maximum-spread} was given. Moreover,
several properties of PPs and effective measures that are relevant for turbo 
coding were investigated. We proposed a new {\em refined product measure}
$\Omega^\prime$ for PP interleavers that is easily and quickly
computed; a table of good interleavers according to the new measure is
provided for several of the interleaver lengths reported in the
literature. QPPs are convenient because of their small
footprint; however, we provided some evidence that PPs with larger degrees
than two may be necessary for very low FERs. Although, turbo codes
are limited by a logarithmic growth on the minimum
distance~\cite{breiling}, that result is asymptotic: good codes at
practical lengths achieving FERs close to $10^{-8}$ without
signs of an error floor have been demonstrated
earlier~\cite{crozier_MPEG} using DRP interleavers. Similar results
are demonstrated in this paper with QPP interleavers. 

The work in~\cite{sun:tak:pp} opened the doors 
to a class of interleavers for turbo coding that provides
the combination of an excellent performance, elegant algebraic and
geometric analysis, and simplicity of implementation; to the best of
our knowledge, there are no other interleavers sharing simultaneously these
three characteristics to their fullest extent. Many problems in turbo
coding that were earlier treated  empirically may now be recast to be
solved systematically and rigorously: the design of multiple turbo
codes~\cite{div:pol:tda42-121} for an improved asymptotic minimum 
distance by classifying the properties of PP interleavers under
function composition; and  puncturing of code bits for higher data
rates using the $MCF$ theory~\cite{tak:mcf} (a stronger generalization
of an even-odd interleaver~\cite{barbulescu}) are only a couple of
relevant examples.  

\begin{center}
  {\bf APPENDIX}
\end{center}
\setcounter{section}{0}
\section{New upper bound on the spread factor $D_E$}
\label{sec:bound}
\begin{figure}[htbp]
  \centering
  \includegraphics[width=0.5\hsize]{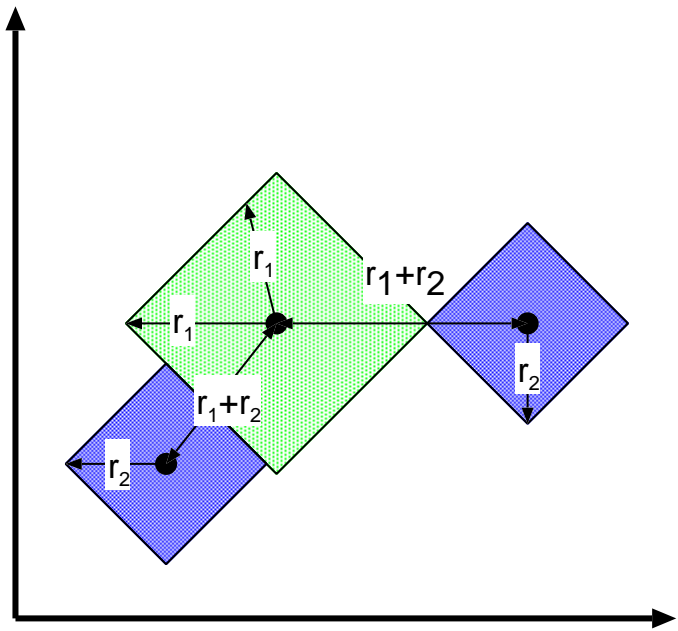}
  \caption{Spheres in the metric space $({\cal V},\delta)$.}
  \label{fig:tspace}
\end{figure}

A technique inspired from~\cite{boutillon} is used to compute the upper bound
$ub_{D_E}(N)$. We use the same distance $\delta$ in
Section~\ref{sec:ms}
but over the space $[0,N-1)^2\subset \mathbb{R}^2$ and
forming the metric space ${\cal V}=([0,N-1)^2,\delta)$.
In this metric space, a sphere of
radius $r_1$ has the shape of a square when projected
over the Euclidean metric space  as shown in
Fig.~\ref{fig:tspace}. If two spheres of radii $r_1$ and $r_2$ touch
each other then the distance between their centers is $r_1+r_2$.

\begin{figure}[htbp]
  \centering
  \includegraphics[width=0.5\hsize,angle=270]{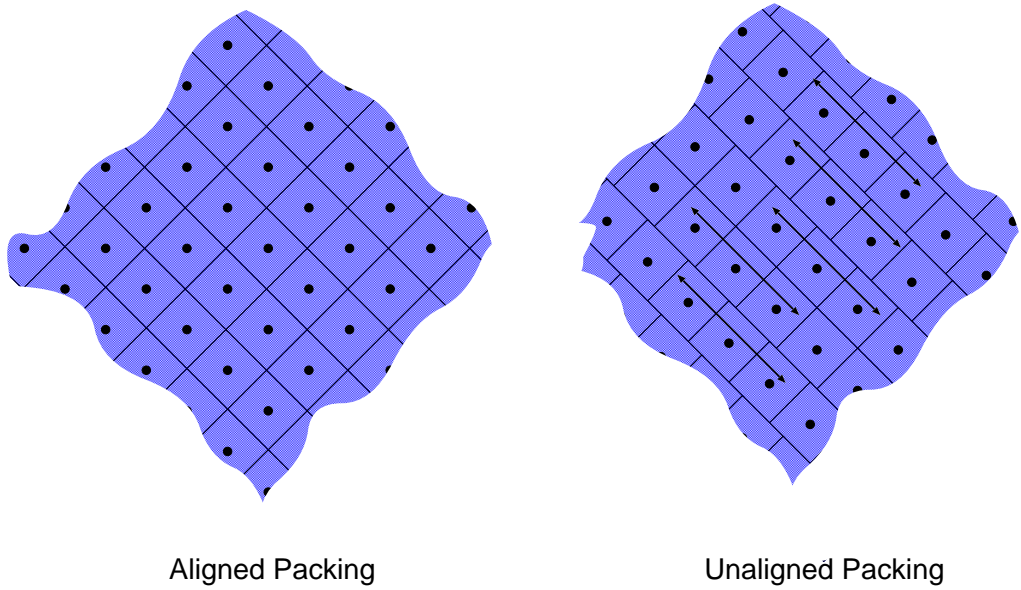}
  \caption{Sphere packing over the metric space $(\mathbb{R}^2,\delta)$.}
  \label{fig:tpacking}
\end{figure}

Before proceeding with the upper bound, it is straightforward to
realize that in the metric space $(\mathbb{R}^2,\delta)$ there are only two
types of densest packings for spheres with identical radius: aligned
packing and unaligned packing as shown in Fig.~\ref{fig:tpacking}. The
unaligned packing is obtained by arbitrarily ``sliding'' stripes of
the aligned packing.  

\begin{figure}[htbp]
  \centering
  \includegraphics[width=0.7\hsize]{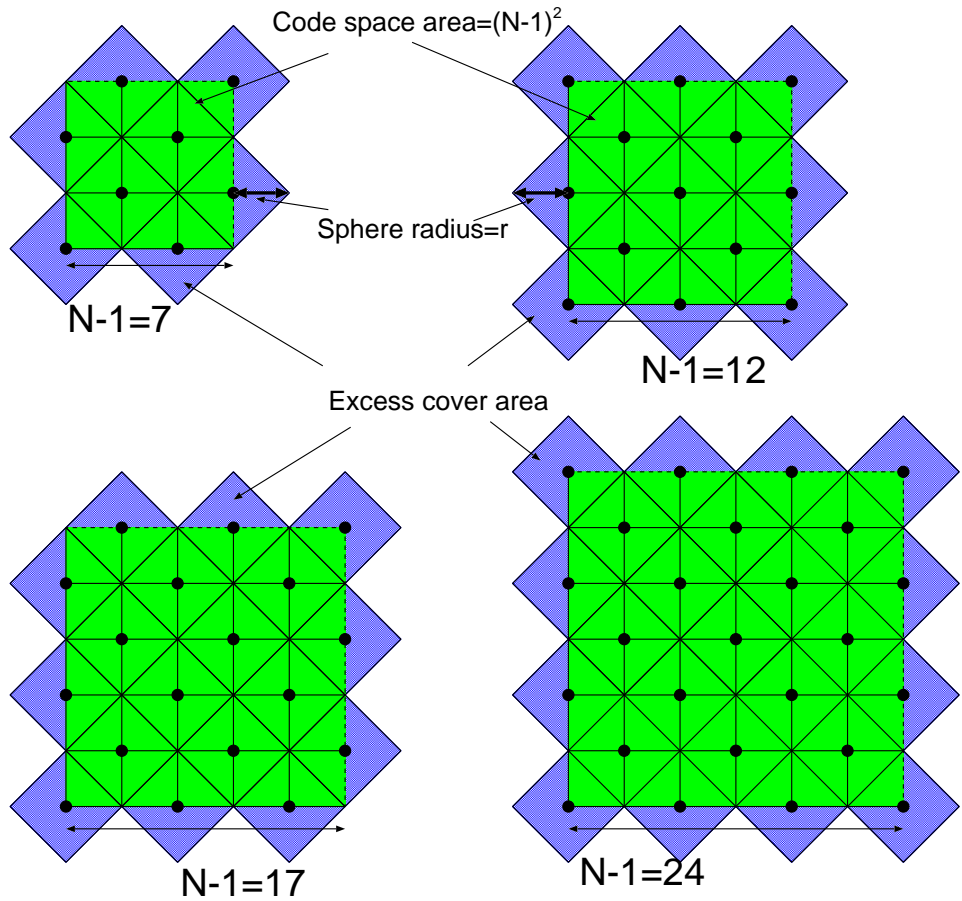}
  \caption{Sphere packing over the metric space $({\cal V},\delta)$.}
  \label{fig:tbound}
\end{figure}

The upper bound on $D_E$ is constructive. Given a square whose area is
$(N-1)^2$, we attempt to pack $N$ spheres of identical radii $r$ and
maximize the radii of the spheres $r$. We do so for values of $N$ for
which clearly a subset of the aligned packing gives the densest
packing as shown in Fig.~\ref{fig:tbound}. We end up with
1) several points touching the boundaries of ${\cal V}$ and 2) 
excess cover areas.

There are two classes of values for $N$. The cases
$N=8$ and $N=18$ belong to

\[
N=2p^2, \quad p=2,3,4\ldots.
\]

\noindent
where $p$ is the number of points on the bottom edge. The upper bound
is the distance between adjacent points, i.e.,
$ub_{D_E}(N)=(N-1)/(2p-1)$. After some simple algebra, the upper bound
$ub_{D_E}(N)$ is 

\[
ub_{D_E}(N)=\frac{2(N-1)}{\sqrt{2N}-1}
\]

The cases $N=13$ and $N=25$ belong to 

\[
N=p^2+(p-1)^2, \quad p=2,3,4,\ldots.
\]

The upper bound $ub_{D_E}(N)$ is

\[
ub_{D_E}(N)=\frac{2(N-1)}{\sqrt{2N-1}-1}.
\]

\section{A Geometric Approach for Maximum-Spread Linear
Interleavers}

\begin{figure}[htbp]
  \centering
  \includegraphics[width=0.7\hsize]{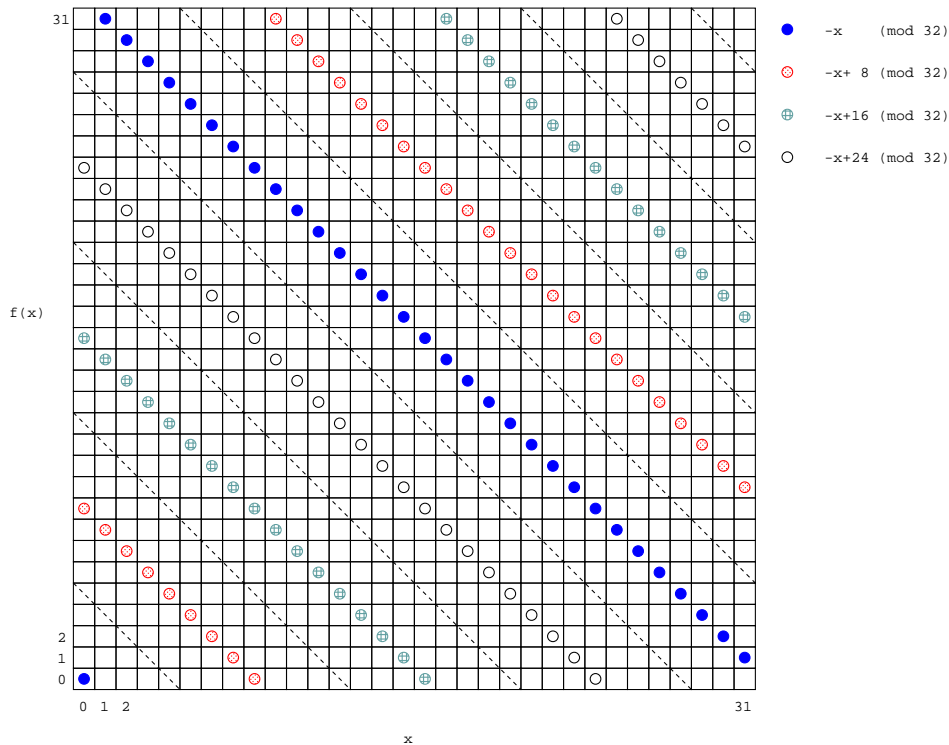}
  \caption{Linear curves describing the ``centers'' of stripes in a
    perfect packing for $N=32$.}
  \label{fig:perfect}
\end{figure}

Although not explained in~\cite{dolinar}, it is clear that many
candidates for an interleaver length $N$ achieving maximum-spread
$\sqrt{2N}$ are of the form $N=2n^2$ because those allow the optimal
packing in $({\cal  V},\delta_N)$. 
This is illustrated in Fig.~\ref{fig:perfect} for $N=32$. The dots
show the potential places that may be occupied by a point of the
interleaver-code. Without loss of generality, we place first a point in
$(0,0)$. Next, we place a point 
along the line $-x+8 \pmod{32}$ and $0<x<8$ with the constraint that a
linear curve that interpolates the first two points is a permutation
polynomial. Then we follow a similar but simpler technique as we do to
prove Theorem~\ref{th:sequence} in Appendix~\ref{sec:sequence}. This
leads to an alternate geometric construction for all maximum-spread
linear interleavers (up to symmetries).   

The authors of~\cite{crozier2} explain the so-called dithered
diagonal construction using the same geometric 
approach. The difference with linear interleavers is that the
placement of a second point does not fix all the remaining
points. Therefore a different slide (or dither) is allowed for the
other stripes provided that an interleaver constraint is maintained. 

\section{Proof of Theorem~\ref{th:sequence}}
\label{sec:sequence}
The set of linear curves $b_i(x)=-x+i2n\pmod{N}, \quad 0\leq i <n$ is simply
passing through the center of the stripes in either the aligned or an unaligned
packing as show in Fig.~\ref{fig:perfect}, i.e., the smallest
distance $\delta_N$ between two points that belong to any two disjoint curves is
$2n$. If we show that the intersection of each of the curves $b_i(x)$  
with $f(x)$ has $N/n$ solutions that are equally spaced by
$\delta_N=2n$ then we are done. 

We start by rewriting the infinite sequence (\ref{eq:sequence}) as
function of $n=2^{k-1}$. 

\begin{equation}
(2^k-1)x+2^{k+1}x^2\equiv (2n-1)x+8nx^2 \pmod{2^{2k-1}=2n^2}
\label{eq:n_form}
\end{equation}

The intersection of (\ref{eq:n_form}) and a linear curve $b_i(x)$ is
the solution to

\[
(2n-1)x+8nx^2\equiv -x+i2n \pmod{2n^2},
\]

or

\begin{equation}
2n(\underbrace{-i+x+4x^2}_{j})\equiv 0 \pmod{2n^2}.
\label{eq:change_var}
\end{equation}

We then transform the quadratic congruence (\ref{eq:change_var}) into a linear congruence

\begin{equation}
2nj\equiv 0 \pmod{2n^2},
\label{eq:lin_reduction}
\end{equation}

\noindent
whose set of solutions is

\[
j\in \{0,n,2n,3n,\ldots,(2n-1)n\},
\]

\noindent
i.e., we have $2n=N/n$ solutions that are equally spaced for
(\ref{eq:lin_reduction}). To solve (\ref{eq:change_var}), we simply
need to find the solution to

\[
-i+x+4x^2\equiv j \pmod{2n^2},
\]

\noindent
when $i$ and $j$ are fixed. Observing that $4x^2+x \pmod{2n^2}$ is a
QPP, the solution for $x$ is unique. Further, from the proof of the
$MCF$ property of permutation polynomials~\cite[Theorem 1]{tak:mcf}, we know
that if $M$ divides $N$ then a set of $M$ equally spaced (modulo
$N$) values by $N/M$ is always mapped to an equally spaced set of
values by a PP and also by its inverse PP. Therefore we establish the
desired result. 

\section{Proof of Theorem~\ref{th:qpp_nl_degree}}
\label{sec:qpp_nl_degree}

If ${\cal A}(k_0,k_1)$ is an isometry of the interleaver-code
$Q=\Phi(q)$  then we must have $q(x-k_0)+k_1\equiv
q(x)\pmod{N}$. Developing it we have

\[
q_2(x-k_0)^2+q_1(x-k_0)+k_1\equiv q_2x^2+q_1x\pmod{N}
\]

\[
q_2x^2+(q_1-2q_2k_0)x-q_1k_0+q_2k_0^2+k_1\equiv q_2x^2+q_1x\pmod{N}
\]

\noindent
and therefore we just need to ensure $2q_2k_0\equiv 0\pmod{N}$. This
is a linear congruence and its solution is completely characterized in
elementary number theory textbooks.

\[
k_0(i)=\frac{Ni}{\gcd(2q_2,N)}, \quad i=0,1,2,...,\gcd(2q_2,N) -1
\]

\noindent
where $k_0(i)$ is the $i$-th solution and

\begin{equation}
k_1(i)=q_1k_0(i)- q_2k_0(i)^2
\end{equation}

\noindent 
There are exactly $\gcd(2q_2,N)$ distinct solutions, which means that
each point in $Q$ belongs to an orbit of size $\gcd(2q_2,N)$, i.e., the
degree of shift-invariance is $\epsilon(Q)=\gcd(2q_2,N)$. Hence, from
Proposition~\ref{prop:shift-nl}, the degree of non-linearity for QPPs is
$\zeta(Q)=N/\gcd(2q_2,N)$.

\end{document}